\documentclass[aps,prd,superscriptaddress,nofootinbib,11pt]{revtex4}
\usepackage[english]{babel}
\usepackage[utf8]{inputenc}
\usepackage{graphicx}   
\usepackage{slashed}
\usepackage{epstopdf}
\usepackage{verbatim}   
\usepackage{color}      
\usepackage{subfigure}  
\usepackage{multirow}
\usepackage{hyperref}   
\usepackage{float}
\usepackage{epsfig,rotating}
\usepackage{amsmath,amssymb}
\usepackage{dsfont}
\restylefloat{table}
\numberwithin{equation}{section}

\newcommand{\vx}{\vec{x}}
\newcommand{\vp}{\vec{p}}
\newcommand{\vq}{\vec{q}}
\newcommand{\vk}{\vec{k}}

\newcommand{\be}{\begin{equation}}
\newcommand{\ee}{\end{equation}}
\newcommand{\bea}{\begin{eqnarray}}
\newcommand{\eea}{\end{eqnarray}}

\newcommand{\ket}[1]{|#1\rangle}
\newcommand{\bra}[1]{\langle#1|}
\newcommand{\order}[1]{\mathcal{O}(#1)}

\begin{document}
\title{Ultralight dark matter or dark radiation cosmologically produced from infrared dressing.}
\author{Daniel Boyanovsky}
\email{boyan@pitt.edu} \affiliation{Department of Physics and
Astronomy, University of Pittsburgh, Pittsburgh, PA 15260.}
\author{Mudit Rai}
\email{MUR4@pitt.edu} \affiliation{Department of Physics and
Astronomy, University of Pittsburgh, Pittsburgh, PA 15260.}
\author{Lisong Chen}
\email{lic114@pitt.edu} \affiliation{Department of Physics and
Astronomy, University of Pittsburgh, Pittsburgh, PA 15260.}

 \date{\today}

\begin{abstract}
Infrared dressing of    bosonic or fermionic heavy particles by a cloud of massless particles to which they couple is studied as a possible production mechanism of ultra light dark matter or dark radiation in a radiation dominated cosmology. We implement an adiabatic expansion valid for wavelengths much smaller than the Hubble radius combined with a non-perturbative and manifestly unitary dynamical resummation method to study the time evolution of an initial single heavy particle state. We find a striking resemblance to the process of particle decay: the initial amplitude of the single particle decays in time, not exponentially but with a power law with anomalous dimension $\propto t^{-\Delta/2}$ featuring a crossover to $t^{-\Delta}$ as the heavy particle becomes non-relativistic  in both bosonic and fermionic cases suggesting certain universality. At long time the asymptotic state is an entangled state of the heavy and massless particles. The entanglement entropy is shown to grow under time evolution describing the flow of information from the initial single particle to the final multiparticle state. The expectation value of the energy momentum tensor in the asymptotic state is described by two indpendent fluids each obeying covariant conservation, one of heavy particles and the other of relativistic (massless) particles (dark radiation). Both fluids share the same frozen distribution function and entropy as a consequence of entanglement.

\end{abstract}


\maketitle
\tableofcontents

\section{Introduction}

Light and ultra-light  particles in extensions beyond the Standard model, such as  axions or axion-like particles, ``fuzzy'' dark matter (FDM), light dark scalars\cite{axionreviu}-\cite{dmpart} could   be   suitable cold-dark matter (CDM) candidates, and dark vector bosons may contribute to a dark radiation component. An (FDM) particle with mass $m\simeq 10^{-22}\,\mathrm{eV}$ has the potential for solving some small scale aspects of galaxy formation\cite{fuzzyDM}-\cite{jens}.

  All of these candidates are characterized by very small masses and couplings to Standard Model degrees of freedom or beyond. Lyman-$\alpha$\cite{Lyalfa,Lyalfaforest} and pulsar timing\cite{pulsarultralight} provide constraints on the mass range of (ultra) light dark matter (ULDM). Light dark matter (DM) candidates are not only probed by their gravitational properties\cite{DMgraviprobes} but there are various proposals for  direct detection,  from high energy colliders\cite{HECdarkfoton} to ``table-top'' experiments\cite{admx}-\cite{savas}. Various possible
  mechanisms of production of light or ultra-light dark matter have
  been discussed in the literature\cite{axionreviu}-\cite{dmpart} including non-adiabatic gravitational production\cite{herruldm}.

  In this article we explore the dynamics of infrared dressing in non-gauge theories as a   \emph{possible} non-thermal  cosmological production mechanism of either ultra light dark matter or   dark radiation prior to recombination. Infrared dressing refers to the cloud of nearly on-shell massless quanta that dresses the charged particle to which these massless fields couple. Infrared singularities associated with the emission and/or absorption of soft massless   quanta by charged fields continues to be the subject of study within the context of  the S-matrix in gauge theories\cite{bn,lee,chung,kino,kibble,yennie,weinberg,kulish,lavelle} and more generally  of infrared phenomena\cite{zell,tomaras,schwartz1,finites,schwartz2}, including in gravity where the emission and absorption of gravitons yields similar infrared effects\cite{strominger1,strominger2}.

  \vspace{1mm}

  \textbf{Motivations and objectives:}
  A recent study\cite{infrared} of the dynamics of infrared dressing from the emission and absorption of massless quanta by a charged massive particle in Minkowski space-time has revealed a striking resemblance to a decay process: the amplitude of the initial state decays in time, yielding as an asymptotic state an entangled state of the massive and massless particles. Although the decay in time of the amplitude of the initial state is not exponential    but as a power law   with an anomalous dimension\cite{infrared},   the asymptotic quantum state  is, in fact, qualitatively similar: a quantum state in which the daughter particles are kinematically entangled\cite{boyww}. This similarity suggests that just as in particle decay,   infrared dressing is an effective production mechanism of (nearly) massless particles.

   Motivated by these results in Minkowski space-time, our objectives in this study are twofold:

   \textbf{a:)} to study   the dynamics  of infrared dressing as a fundamental process in a radiation dominated (RD) cosmology, with direct relevance in gauge theories and gravity.

   \textbf{b:)} A ``\emph{proof of principle}'' of infrared dressing as a \emph{possible} non-thermal production mechanism of ultra light dark matter or dark radiation prior to recombination.

    Neither of these aspects of infrared phenomena has been hitherto addressed in cosmology, thus this study represents a first step towards a more comprehensive treatment of   these phenomena in connection with ultra light dark matter and/or dark radiation in extensions beyond the Standard Model. Ultra light dark matter or dark radiation with a non-thermal frozen distribution may contribute to the effective number of ultrarelativistic species prior to recombination, but their contribution depends crucially on their non-thermal distribution\cite{planck}, which in turn depends on the dynamics of the production mechanism. Therefore, the study of infrared dressing in cosmology  may reveal a new
a production mechanism leading to a non-thermal frozen distribution that may evade current cosmological bounds\cite{bounds}.
  \vspace{1mm}

  \textbf{Brief summary of results:}
  In this article we focus on the dynamics of infrared dressing of a \emph{single} heavy particle resulting from the emission/absorption of soft quanta in a radiation dominated cosmology in non-gauge theories, as a prelude towards a study   of an ensemble of heavy particles during this era.

    We implement the dynamical resummation method (DRM) introduced in refs.\cite{infrared,decaycosmo} (and references therein) combined with an adiabatic expansion valid for wavelengths much smaller than the particle horizon\cite{decaycosmo,scattering}, to study the infrared dressing of heavy particles by soft quanta of a massless (or nearly massless) scalar field in a radiation dominated cosmology. While we focus on non-gauge theories, thereby  bypassing the important and subtle  issue of gauge invariance, postponed to a future study, the nature of the infrared divergences in the  case of massless scalar fields is akin to those in gauge theories\cite{infrared}.

    Two models are considered:  a heavy complex scalar minimally coupled to gravity and coupled to a massless scalar field and a heavy fermion Yukawa coupled to the massless scalar field. The massless scalar field is  taken to possibly describe the ultra-light dark matter or dark radiation particle, which  could be a (pseudo) Goldstone boson in a suitable extension beyond the Standard Model. We do not specify nor address the nature or phenomenology of this field since our main objective is to focus on the fundamental aspects and a proof of principle of the production mechanism.

    Our study shows that infrared dressing is qualitatively similar to particle decay in that the amplitude of the initial single particle state decays in time, not as an exponential modified by the expansion\cite{decaycosmo} as in the case of particle decay, but as a power law with an anomalous dimension $\propto [E_k(t)t]^{-\Delta}$ with $E_k(t)$ the local energy of the heavy particle. Bosonic and fermionic heavy particles feature the same long time behavior with different anomalous dimensions suggesting an universality for infrared phenomena in cosmology.  At long time the asymptotic state is an entangled state of the heavy and massless particles with the total probability of this entangled state saturating  the unitarity condition. Entanglement of the asymptotic state is confirmed by obtaining the entanglement (von Neumann) entropy, which describes the information flow from the initial single particle to the asymptotic multiparticle state. The entanglement entropy is shown to increase in time and its time evolution   is completely determined by the (DRM) equations.  The expectation value of the energy momentum tensor in the asymptotic state describes two independent fluids each   satisfying covariant conservation, one associated with the heavy particle and another describing a relativistic particle associated with either ultra light dark matter or dark radiation. Entanglement in the asymptotic state results in that both fluids share the same frozen distribution function and entropy.

    The article is organized as follows: in section (\ref{sec:bosonic}) we consider a bosonic model of a heavy boson interacting with a massless boson. We discuss  its quantization aspects and introduce the adiabatic expansion in detail explaining its physical underpinning. In section (\ref{sec:DRM}) we introduce the dynamical resummation method (DRM) described in ref.\cite{infrared} extended to cosmology in conjunction with the adiabatic expansion and apply it to the bosonic model. We show that an initial single heavy particle state decays in time as $[E_k(t)t]^{-\Delta}$ with $E_k(t)$ the local energy of the heavy particle and $\Delta$ an anomalous dimension, displaying a crossover from $\propto t^{-\Delta/2}$ early when it is relativistic to $\propto t^{-\Delta}$ when it becomes non-relativistic. We extract  the asymptotic state obtained from the relaxation of an  initial single heavy particle state and show that it is an entangled state of the heavy and the massless bosons with a frozen non-thermal distribution function.

In section (\ref{sec:fermions}) we consider a heavy fermion Yukawa coupled to a massless scalar, quantizing the theory, introducing the adiabatic expansion   and the dynamical resummation method for  fermions. We show that   the amplitude of an initial single particle heavy fermion state decays in time in a manner qualitatively similar  to the bosonic case, indicating certain universality in the cosmological dynamics of infrared dressing\cite{infrared}. The asymptotic state is, again, an entangled state of the heavy fermion and the massless particle with a frozen non-thermal distribution.

In section (\ref{sec:emt}) we study the energy momentum tensor in the asymptotic regime when the amplitude of initial states is vanishingly small in both cases. Entanglement between the heavy and massless degrees of freedom in the asymptotic state is confirmed by obtaining the (entanglement) von Neumann entropy by tracing either one of the degrees of freedom. To leading order in the adiabatic expansion and couplings, the energy momentum tensor describes two independent fluids: one of massive particles and another of radiation, both are determined by the non-thermal frozen distribution associated with the asymptotic entangled state, and share the same distribution function and entropy as a consequence of entanglement.

Section (\ref{sec:discussion}) discusses various aspects and caveats of the results, conclusions and further questions are summarized in section (\ref{sec:conclusions}).

\section{Bosonic case: quantization and adiabatic expansion}\label{sec:bosonic}

We focus our study on the infrared dynamics in a spatially flat  Friedmann-Robertson-Walker (FRW) cosmology.     In conformal time $\eta$  with $d\eta = dt/a(t)$, the metric is given by
\be  g_{\mu \nu} = a^2(\eta)\,\textrm{diag}(1,-1,-1,-1) \, . \label{conformalmetric} \ee

In the standard cosmological picture  most of the interesting particle physics processes occur during the RD era and   we focus   our attention on this epoch, during which the scale factor in conformal time is given by
\be a(\eta)= H_R   ~\eta  ~~;~~ H_R= H_0\,\sqrt{\Omega_R}\simeq 10^{-44}\,\mathrm{GeV}\,.  \label{aofRD}\ee In a radiation dominated cosmology the Ricci scalar vanishes, therefore massless particles are conformally coupled to gravity during this epoch.

 During the (RD) stage   the relation between conformal and comoving time is given by \be \eta = \Big( \frac{2\,t}{H_R}\Big)^{\frac{1}{2}} \Rightarrow a(t) = \Big[ 2\,t H_R\Big]^{\frac{1}{2}}\,, \label{etaoft} \ee a result that will prove useful in the study of the (comoving) time dependence of amplitudes during this stage.

We begin by considering the simpler case of   two interacting scalar fields minimally coupled to gravity, a massive complex (charged) field $\Phi$ and a massless neutral field $\pi$,   with action given by
\be  A   =    \int d^4 x \sqrt{|g|} \Bigg\{  g^{\mu\nu}\,\partial_\mu \Phi^\dagger \partial_\nu \Phi-  M^2\,\Phi^\dagger\,\Phi + \frac{1}{2} g^{\mu\nu}\,\partial_\mu \pi \partial_\nu \pi  -   \lambda\, :\Phi^\dagger \,\Phi: \,  \pi  \Bigg\}
\label{action}\ee   where normal ordering is understood in the interaction picture of free fields.

Expressing the action of Eq.~(\ref{action}) in terms of   comoving spatial coordinates and   conformal time, and  conformally rescaling the fields as
\be \Phi(\vec{x},t) = \frac{\varphi(\vec{x},\eta)}{a(\eta)} ~~;~~ \pi(\vec{x},t) = \frac{\chi(\vec{x},\eta)}{a(\eta)}  \,, \label{rescale}\ee
yields
\be A= \int d^3x \, d\eta  \biggl\{   \frac{d\varphi^\dagger}{d\eta}\, \frac{d\varphi}{d\eta}  -  \nabla \varphi^\dagger \cdot \nabla \varphi  - M^2\,a^2(\eta)\,\varphi^\dagger\,\varphi + \frac{1}{2}
\Big(\frac{d\chi}{d\eta}\Big)^2   -  \frac{1}{2}\Big(\nabla \chi \Big)^2  - \lambda \,a(\eta)\,:\varphi^\dagger\,\varphi: \chi  \biggr\}\,
\label{conformalaction}\ee where, as usual, we have neglected total surface terms which do not contribute to the equations of motion.

\vspace{2mm}

  We begin with the quantization of free fields \cite{parker,ford,zelstaro,birrell,fullbook,parkerbook,mukhabook,birford,bunch,parfull}   as a prelude to the interacting theory. The Heisenberg equations of motion for the conformally rescaled fields $\varphi, \chi$ in conformal time are
\bea \frac{d^2}{d\eta^2}\,\varphi(\vec{x},\eta) - \nabla^2 \varphi(\vec{x},\eta) + M^2\,a^2(\eta)\,\phi(\vec{x},\eta) &  = & 0 \,,\label{Eomvarphi}\\\frac{d^2}{d\eta^2}\,\chi(\vec{x},\eta) - \nabla^2 \chi(\vec{x},\eta)   &  = & 0 \,. \label{Eomchi}
\eea
It is convenient to quantize the fields  in a comoving volume $V$, in a plane wave expansion in terms of comoving wave vectors $\vec{k}$ and  comoving coordinates $\vec{x}$,  namely,
\be   \varphi(\vx,\eta) =  \frac{1}{\sqrt{ V}}\, \sum_{\vk}\Big[ a_{\vk} \,g_k(\eta)\, e^{i\vk\cdot\vx} + b^\dagger_{\vk} \,g^*_k(\eta) \, e^{-i\vk\cdot\vx}\Big]\,. \label{fige} \ee
\be   \chi(\vx,\eta) =  \frac{1}{\sqrt{ V}}\, \sum_{\vk}\Big[ c_{\vk} \,f_k(\eta)\, e^{i\vk\cdot\vx} + c^\dagger_{\vk} \,f^*_k(\eta) \, e^{-i\vk\cdot\vx}\Big]\,,  \label{chief} \ee
 where the mode functions $g_k(\eta);f_k(\eta)$ are solutions of the following equations
 \bea \Big[ \frac{d^2}{d\eta^2}+ \Omega^2_k(\eta)   \Big]g_k(\eta)  & = & 0 ~~;~~ \Omega^2_k(\eta) = k^2 + M^2\,a^2(\eta) \label{equvarfi} \\
 \Big[ \frac{d^2}{d\eta^2}+ k^2  \Big]f_k(\eta)  & = & 0    \,, \label{equchi}\eea  and satisfy   the Wronskian condition
\begin{align}
& g^{\,'}_k(\eta)g^*_k(\eta) - g^{*\,'}_k(\eta) g_k(\eta) = -i \,  \\
& f^{\,'}_k(\eta)f^*_k(\eta) - f^{*\,'}_k(\eta) f_k(\eta) = -i \,, \label{wronski}
\end{align}
so that the annihilation   and creation   operators are \emph{time independent} and obey the canonical commutation relations $[a_{\vk},a^{\dagger}_{\vk'}] = \delta_{\vk,\vk'} \,; [c_{\vk},c^{\dagger}_{\vk'}] = \delta_{\vk,\vk'}\,\mathrm{etc.} $.
The vacuum state $\ket{0^{\varphi};0^{\chi}}$ is defined such that
\be a_{\vk}\ket{0^{\varphi};0^{\chi}}= b_{\vk}\ket{0^{\varphi};0^{\chi}} = c_{\vk}\ket{0^{\varphi};0^{\chi}} =0 \,. \label{vacuumdef}\ee

The mode functions $f_k(\eta)$ solutions of eqn. (\ref{equchi}) obeying (\ref{wronski}) are given by
\be f_k(\eta)= \frac{e^{-ik\eta}}{\sqrt{2k}}\,. \label{fofk}\ee
Introducing the dimensionless variables
\be x = \sqrt{2\,M\,H_R} ~ \eta ~~;~~ \alpha = -\frac{k^2}{2\,M\,H_R}\,, \label{weberparas} \ee in terms of which the equation (\ref{equvarfi}) is identified with Weber's equation\cite{gr,as,nist,bateman,magnus}
\be \frac{d^2}{dx^2}\,w(x) +\Big[\frac{x^2}{4}-\alpha \Big]w(x) =0 \,.\label{webereq}\ee The general solutions are  linear combinations of Weber's parabolic cylinder functions $W[\alpha;\pm x]$\cite{gr,as,nist,bateman,magnus}. These are real solutions, hence we seek a linear combination  that can be identified with    particle states asymptotically at long time.

To understand the asymptotic behavior at long time  we will carry out a  Wentzel-Kramers-Brillouin (WKB) analysis for $g_k(\eta)$. Writing the solution of the mode equations (\ref{equvarfi}) in the WKB form\cite{birrell,fullbook,mukhabook,parkerbook,birford,bunch,mottola,dunne,wini}
\be g_k (\eta) = \frac{e^{-i\,\int^{\eta}_{\eta_i}\,W_k(\eta')\,d\eta'}}{\sqrt{2\,W_k(\eta)}} \,, \label{WKB}\ee and inserting this ansatz into (\ref{equvarfi}) it follows that $W_k(\eta)$ must be a solution of the equation\cite{birrell}
\be W^2_k(\eta)= \Omega^2_k(\eta)- \frac{1}{2}\bigg[\frac{W^{''}_k(\eta)}{W_k(\eta)} - \frac{3}{2}\,\bigg(\frac{W^{'}_k(\eta)}{W_k(\eta)}\bigg)^2 \bigg]\,.  \label{WKBsol} \ee

This equation can be solved in an \emph{adiabatic expansion}
\be W^2_k(\eta)= \Omega^2_k(\eta) \,\bigg[1 - \frac{1}{2}\,\frac{\Omega^{''}_k(\eta)}{\Omega^3_k(\eta)}+
\frac{3}{4}\,\bigg( \frac{\Omega^{'}_k(\eta)}{\Omega^2_k(\eta)}\bigg)^2 +\cdots  \bigg] ~~;~~ \Omega_k(\eta) = \sqrt{k^2+M^2a^2(\eta)} \,.\label{adexp}\ee We refer to terms that feature $n$-conformal time derivatives of $\Omega_k(\eta)$ as of n-th adiabatic order. The nature and reliability of the adiabatic expansion is revealed by considering the term of first adiabatic order, namely:
\be \frac{\Omega^{'}_k(\eta)}{\Omega^2_k(\eta)} = \frac{M^2\, a(\eta) a^{'}(\eta)}{\Big[ k^2 + M^2\,a^2(\eta) \Big]^{3/2}}\,, \label{firstordad}\ee this is most easily recognized in \emph{comoving} time $t$, introducing the \emph{local} energy $E_k(t)$ and Lorentz factor $\gamma_k(t)$ measured by a comoving observer in terms of the \emph{physical} momentum $k_p(t) = k/a(t)$
\bea E_k(t) & = &  \sqrt{k^2_p(t)+M^2} = \frac{\Omega_k(\eta)}{a(\eta)}  \label{comoener}\\
 \gamma_k(t) & = & \frac{E_k(t)}{M} \,,\label{gamafac} \eea  and the Hubble expansion rate
 $H(t) = \frac{\dot{a}(t)}{a(t)} = a^{'}/a^2 $.    In terms of these variables, the first order adiabatic ratio  (\ref{firstordad}) becomes
 \be \frac{\Omega^{'}_k(\eta)}{\Omega^2_k(\eta)} = \frac{H(t)}{\gamma^2_k(t)\,E_k(t)}\,.  \label{adratio}\ee

 In similar fashion the higher order terms in the adiabatic expansion for a (RD) cosmology (vanishing Ricci scalar) can be obtained,
\begin{align}
\frac{\Omega^{''}_k(\eta)}{\Omega_k^3(\eta)} &= \frac{1}{\gamma^2_k(t)} \frac{H^2(t)}{E_k^2(t)}\Big[1-\frac{1}{\gamma^2_k(t)}\Big] \nonumber\\
\frac{\Omega^{'''}_k(\eta)}{\Omega_k^4(\eta)} &= -  \frac{3}{\gamma^3_k(t)}\,\frac{H^3}{E^3_k} \Big[1-\frac{1}{\gamma^2_k(t)} \Big] \,. \label{secthirdad}
\end{align}
   Consequently, (\ref{adexp}) takes the form:
\begin{equation}
W^2_k(t) = a^2(t)E^2_k(t)\Big[1-\frac{1}{2\gamma^2_k(t)} \frac{H^2(t)}{E_k^2(t)}\Big[ 1-\frac{5}{2\gamma^2_k(t)}\Big] +\cdots \Big]\,.
\end{equation}

From the above analysis it is clear that
\be  \frac{H(t)}{\gamma_k(t)\,E_k(t)} \ll 1 \,, \label{adpara}\ee is the small, dimensionless \emph{adiabatic} expansion parameter.  We will instead adopt a more stringent condition for validity of the adiabatic approximation, namely
\be \frac{H(t)}{E_k(t)} \ll 1 \Rightarrow E_k(t)\,t \gg 1\,,  \label{adiarat} \ee where we used the relation (\ref{etaoft}) in the second inequality.

  The physical interpretation of the   ratio $H(t)/E_k(t)$ is clear: typical particle physics degrees of freedom feature either physical de Broglie or Compton wavelengths that are much smaller than the (physical) particle horizon (or Hubble radius) $\propto 1/H(t)$   at any given time during (RD).

  In a standard (RD) cosmology the particle horizon always grows faster than a physical wavelength, therefore the reliability of the adiabatic expansion improves with the cosmological expansion. The condition (\ref{adiarat}) is also equivalent to a ``long time limit'' in the sense that there are many oscillations of the microscopic degrees of freedom within a Hubble time $\simeq 1/H(t)$.

  Therefore the validity  of the adiabatic expansion hinges on the separation of the two relevant time scales: the slow time scale of cosmological expansion $\simeq 1/H(t)$ and the rapid time scale associated with the oscillations of the field $\simeq 1/E_k(t)$, with $E_k(t)/H(t) \gg 1$\cite{scattering}.

  In an (RD) cosmology with scale factor given by (\ref{aofRD}), it follows that in the adiabatic regime
  \be \frac{\Omega'_k(\eta)}{\Omega^2_k(\eta)}\equiv \widetilde{\epsilon}_k(\eta) =  \frac{\epsilon_k(\eta)}{\gamma^2_k(\eta)}~~;~~ \epsilon_k(\eta) \equiv \frac{H(t)}{E_k(t)} = \frac{1}{\Omega_k(\eta)\,\eta} \ll 1 \,, \label{epsi}\ee where we introduced the small dimensionless parameter $\epsilon_k(\eta)$ that characterizes the adiabatic expansion.   For the purpose of analyzing contributions in the adiabatic expansion, we will consider $\epsilon_k(\eta)$ and $\widetilde{\epsilon}_k(\eta)$ to be of the same order.
  Therefore the  adiabatic expansion is an expansion in the small dimensionless ratio $\epsilon_k(\eta)$ which becomes smaller   upon cosmological expansion.

  Since the adiabatic approximation improves with cosmological expansion, either the short wavelength or the long time limits of the WKB solution (\ref{WKB}) is given by
  \be g_k(\eta) \rightarrow  \frac{e^{-i\,\int^{\eta}_{\eta_i}\,\Omega_k(\eta')\,d\eta'}}{\sqrt{2\,\Omega_k(\eta)}}\,, \label{zerog}\ee which is the zeroth order approximation in the adiabatic expansion. The lower limit $\eta_i$ corresponds to an initial time at which the adiabatic approximation becomes reliable.

  The phase of the mode function has an immediate interpretation in terms of comoving time and the local comoving energy (\ref{comoener}), namely
\be e^{-i\,\int^{\eta}_{\eta_i}\,\Omega_k(\eta')\,d\eta'} = e^{-i\,\int^{t}_{t_i}\,E_k(t')\,dt'}\,.\label{phase}\ee
where we used the relations $\Omega_k(\eta) = a(\eta) E_k(t) ~;~ a(\eta)d\eta = dt$.   This  is a natural and straightforward generalization of the phase of \emph{positive frequency} particle  boundary conditions on the mode functions\cite{herringdm}.

To understand better the nature of the zeroth adiabatic order (\ref{zerog}) let us consider a short time interval in the phase in (\ref{phase}). Writing $E_k(t') \simeq E_k(t_i) - k_{ph}(t_i)\,\beta_k(t_i)\,H(t_i)\,(t'-t_i) + \mathcal{O}((t-t_i)^2+\cdots$ the phase becomes
\be \int^{t}_{t_i}\,E_k(t')\,dt'= E_k(t_i)(t-t_i) \,\Big[ 1- \frac{1}{2}\,\beta^2_k(k_i)\,H(t_i)\,(t-t_i)+ \cdots \Big]~~;~~ \beta_k(t) = \frac{k_{p}(t)}{E_k(t)}\,, \label{minkl}\ee therefore the phase coincides with that expected in Minkowski space-time when $(t-t_i) \ll 1/H(t_i)$, namely when the time scale involved is much smaller than the Hubble time. This is the equivalence principle at work. However, early during the (RD) era, and for processes that occur over long periods of
 time during the expansion history as could be the case for very weakly coupled theories, the full time integral must be considered as it includes memory of  this history.

As an example to clarify the regime of validity of the adiabatic approximation, let us consider processes occurring early  in the (RD) stage, for example   at the Grand Unification   scale  $\simeq 10^{15}\,\mathrm{GeV}$, assuming that particles feature  \emph{physical} momenta at this scale $k_{ph}(\eta) = k/a(\eta) \simeq 10^{15}\,\mathrm{GeV}$ with $k$ being the \emph{comoving} momentum and a mass $\simeq 100\,\mathrm{GeV}$, hence a local Lorentz factor $\gamma_k \simeq 10^{13}$. If the environmental temperature of the plasma is $T \simeq T_{\text{GUT}} \simeq 10^{15}\,\mathrm{GeV}$ and taking as an example the standard model result $g_{eff} \simeq 100$, it follows that $H \simeq 10^{12}\,\mathrm{GeV}$. Approximating
$T_{\text{GUT}} \simeq T_\text{CMB}/a(\eta_i)$, where $T_{CMB}$ is the temperature of the cosmic microwave background today,  implies that the scale factor at the GUT scale $a(\eta_i) \simeq 10^{-28}$  and a  \emph{comoving} wavevector    $k \simeq 10^{-13}\,\mathrm{GeV}$ (the average momentum of a microwave photon today). This situation yields  $\epsilon_k =H/E_k \simeq 10^{-3}$, which becomes smaller with the cosmological expansion and the adiabatic ratio $\widetilde{\epsilon}_k(\eta)$ is even much smaller on  account of the Lorentz factor.   It is the wide separation between the slow Hubble time scale $\propto 1/H(t)$ and the fast oscillation time scale $\propto 1/E_k(t)$  that warrants the adiabatic approximation implemented in  our analysis below.

The exact solution of the mode equations (\ref{equvarfi}) that feature asymptotic \emph{positive frequency} particle  boundary conditions
  \be g_k(\eta) \rightarrow  \frac{e^{-i\,\int^{\eta}_{\eta_i}\,\Omega_k(\eta')\,d\eta'}}{\sqrt{2\,\Omega_k(\eta)}}\,, \label{outstates}\ee  and satisfy the Wronskian condition (\ref{wronski}) were found in ref.\cite{herringdm}, these are given by
\be g_k(\eta) = \frac{1}{(8\,M\,H_R)^{1/4}}\,\Big[\frac{1}{\sqrt{\kappa}}\,W[\alpha;x] -i\sqrt{\kappa}\,W[\alpha;-x]  \Big]~~;~~ \kappa = \sqrt{1+e^{-2\pi|\alpha|}}-e^{-\pi|\alpha|}\,.\label{gconf}\ee It is shown in ref.\cite{herringdm} that the asymptotic behavior of $g_k(\eta)$   is indeed given by (\ref{outstates}) both at  long time and also for large (comoving) wavevectors, or short distance.

In the presence of interactions, obtaining transition matrix elements with the exact mode functions (\ref{gconf}) is a daunting task. To make progress, we will restrict our study by considering (comoving) wavevectors and mass for the heavy degrees of freedom for which the adiabatic expansion is reliable, namely for $\epsilon_k(\eta) = H(t)/E_k(t)=1/(\Omega_k(\eta)\eta) \ll 1$ at all times, and keeping only the leading, (zeroth) order in the adiabatic expansion. In this approximation the quantized fields are
\be   \varphi(\vx,\eta) =  \frac{1}{\sqrt{ V}}\, \sum_{\vk}\frac{1}{\sqrt{2\Omega_k(\eta)}}\Big[ a_{\vk} \,e^{-i\,\int^{\eta}_{\eta_i} \,\Omega_k(\eta')\,d\eta'} \, e^{i\vk\cdot\vx} + b^\dagger_{\vk} \,e^{i\,\int^{\eta}_{\eta_i} \,\Omega_k(\eta')\,d\eta'}\, e^{-i\vk\cdot\vx}\Big]\,, \label{figezero} \ee
\be   \chi(\vx,\eta) =  \frac{1}{\sqrt{ V}}\, \sum_{\vk}\frac{1}{\sqrt{2k}}\Big[ c_{\vk} \,e^{-ik\eta}\, e^{i\vk\cdot\vx} + c^\dagger_{\vk} \,e^{ik\eta} \, e^{-i\vk\cdot\vx}\Big]\,,  \label{chiefmink} \ee and the vacuum state $\ket{0^{\varphi};0^{\chi}}$ is annihilated by $a_{\vk},b_{\vk},c_{\vk}$ as per equation (\ref{vacuumdef}).

While the particle interpretation of the quanta of the massless field $\chi$ is clear from the expansion (\ref{chiefmink}), the particle identification in the massive case is confirmed by considering the free field Hamiltonian in the adiabatic approximation\cite{decaycosmo}. The conformal time free field Hamiltonian for the massive field is given by
\begin{equation}
 H_{0\varphi}(\eta) = \int d^3x\,\Big\{ \pi^\dagger\,\pi+\nabla\varphi^\dagger\cdot \nabla\varphi+M^2 a^2(\eta)\,\varphi^\dagger\,\varphi \Big\}~~;~~ \pi \equiv \varphi' \,, \label{varfiHam}
\end{equation}
with equal conformal time canonical commutation relation
\be \Big[\pi(\vec{x},\eta),\varphi(\vec{y},\eta)\Big] = -i \delta^{(3)}(\vec{x}-\vec{y}) \label{ETC}\,,\ee and  similar commutation relations for the neutral massless field.
Using the expansion (\ref{outstates}) and carrying out the spatial integration, we find
\begin{align}
  H_{0\varphi}(\eta)  =  \sum_{\vec{k}}\Bigg\{
    \Big[a^\dagger_{\vec{k}}  a_{\vec{k}}+b_{\vec{k}}b^\dagger_{\vec{k}}\Big] \, \Big[|g'_k|^2+\Omega^2_k(\eta)\,|g_k|^2\Big]
    +\Big(a_{\vec{k}} b_{-\vec{k}}\,\Big[(g'_k)^2+\Omega^2_k(\eta)(g_k)^2\Big] + h.c.\Big)
    \Bigg\}\,. \label{Hphi}
\end{align} Writing $g_k(\eta)$ in the WKB form (\ref{WKB})  it is straightforward to confirm that the terms   $a_{\vec{k}} b_{-\vec{k}}$ in (\ref{Hphi}) are of second and higher adiabatic order\cite{decaycosmo,scattering}. Keeping the leading zeroth adiabatic order, we find
\be H_{0\varphi}(\eta)= \sum_{\vec{k}}
    \Big[a^\dagger_{\vec{k}}  a_{\vec{k}}+b_{\vec{k}}b^\dagger_{\vec{k}}\Big]\,\Omega_k(\eta)\,, \label{H0phi} \ee with
    \be [H_{0\varphi}(\eta),H_{0\varphi}(\eta')] = 0 \,.  \label{hficom}\ee

  Similarly, for the massless fields,
 \be H_{0\chi} = \sum_{\vk} c^\dagger_{\vk}\,c_{\vk}\,k \,,\label{H0chi} \ee where we neglected a zero point contribution. To leading adiabatic order the total free field Hamiltonian is $H_0(\eta) = H_{0\varphi}(\eta)+H_{0\chi}$  which depends explicitly on time through the time dependent frequencies $\Omega_k(\eta)$ for the massive fields.

 The vacuum state   is defined by eqn. (\ref{vacuumdef}) and particle states are, as usual, obtained by applying the creation operators $a^\dagger_{\vk};b^\dagger_{\vk};c^\dagger_{\vk}$ to the vacuum state. These are instantaneous eigenstates of the zeroth adiabatic order Hamiltonian (\ref{hficom}).

 \vspace{1mm}

 \textbf{Dark radiation vs. Ultra light dark matter:}
 We consider the coupling of the massive to a massless field. This massless field could be a Goldstone boson associated with a broken symmetry beyond the standard model and as such could be a candidate for ``dark radiation''. However, an ultra light boson with mass $\simeq 10^{-22} \, \mathrm{eV}$ can be taken as massless during the radiation era with $a(\eta) \leq 10^{-3}$. Consider  for example a  comoving wavector $k\simeq 10^{-24} \,\mathrm{eV}$ corresponding to a de Broglie wavelength $ \simeq \mathrm{kpc}$, the physical wavevector $k_{ph}(\eta) = k/a(\eta)$ is still much larger than the mass of the ultra light scalar during radiation and the contribution of these wavevectors to the energy momentum tensor are strongly suppressed by the phase space factor $\propto k^2$ (see section (\ref{sec:emt})). Therefore, by considering a massless boson coupled to the heavy degrees of freedom we treat dark radiation and an ultra light dark matter candidate on the same footing during the radiation era.

\section{Dynamical resummation method.}\label{sec:DRM}

In this section we adapt the dynamical resummation method developed in ref.\cite{infrared} to the cosmological setting.

In the Schr\"odinger picture, quantum states obey
\begin{equation}
i\frac{d}{d\eta}\ket{\Psi(\eta)} = H(\eta)\ket{\Psi(\eta)}\,,\label{sch}
\end{equation}
where   the total Hamiltonian carries an explicit $\eta$ dependence.  The solution of (\ref{sch}) is given in terms of the unitary time evolution operator $U(\eta,\eta_i)$, namely
$ \ket{\Psi(\eta)} = U(\eta,\eta_i)\ket{\Psi(\eta_i)}$, and  $U(\eta,\eta_i)$ obeys
\begin{equation}
i\frac{d}{d\eta}U(\eta,\eta_i) = H(\eta)U(\eta,\eta_i)~~;~~ U(\eta_i,\eta_i) =1 \,.\label{Uofeta}
\end{equation}
The initial value problem for the time evolution of states will be initialized at a (conformal) time $\eta_i$, with the main assumption that $\Omega_k(\eta_i)\eta_i \gg 1$, to ensure the validity of the adiabatic approximation.
In the interacting theory  $H(\eta) = H_0(\eta)+H_i(\eta)$, where $H_0(\eta)$ is the free field theory Hamiltonian, which to leading adiabatic order is given by $H_{0\varphi}+H_{0\chi}$,  with $H_{0\varphi}$ given by (\ref{Hphi}) and $H_i(\eta)$ the interaction Hamiltonian. In the absence of interactions with $H_i=0$, the time evolution operator of the free field theory $U_0(\eta,\eta_0)$ obeys
\begin{equation}
i\frac{d}{d\eta}U_0(\eta,\eta_i) = H_0(\eta)U_0(\eta,\eta_i),
\quad
-i\frac{d}{d\eta}U^{-1}_0(\eta,\eta_i) = U^{-1}_0(\eta,\eta_i)H_0(\eta),
\quad U(\eta_i,\eta_i)=1 \,, \label{freeH0}
\end{equation}
 to leading order in the adiabatic approximation it is given by
\be U_0(\eta;\eta_i) = e^{-iH_{0\chi}(\eta-\eta_i)}\otimes e^{-i\int^{\eta}_{\eta_i}H_{0\varphi}(\eta')\,d\eta'}\,, \label{U0}\ee as a consequence of (\ref{hficom}).

 It is convenient to pass to the interaction picture, where the operators evolve with the free field Hamiltonian and the states carry the time dependence from the interaction, namely
\be |\Psi_I(\eta)\rangle = U^{-1}_0(\eta,\eta_i)\,|\Psi(\eta)\rangle\,, \label{psiip}\ee and their time evolution is given by
\be |\Psi_I(\eta)\rangle  = U_I(\eta,\eta_i) \,|\Psi_I(\eta_i)\rangle ~~;~~ U_I(\eta,\eta_i) = U^{-1}_0(\eta,\eta_i)\,U(\eta,\eta_i)\,.  \label{evolip}\ee The unitary time evolution operator in the interaction picture $U_I(\eta ,\eta_i)$ obeys
\begin{equation}
i\frac{d}{d\eta}U_I(\eta,\eta_i) = H_I(\eta)U_I(\eta,\eta_i) ~~;~~ H_I(\eta) = U^{-1}_0(\eta,\eta_i)H_i(\eta)U_0(\eta,\eta_i)~~;~~ U_I(\eta_i,\eta_i) =1\,.  \label{Uip}
\end{equation} For the conformal action (\ref{conformalaction}) it follows that
\be H_I(\eta) = \lambda \, a(\eta) \int d^3x ~ \chi (\vx,\eta)\,:\varphi^\dagger(\vx,\eta)\,\varphi(\vx,\eta):\,, \label{HI}\ee where the fields are given by the free field expansion (\ref{figezero},\ref{chiefmink})   and time independent creation and annihilation operators for the respective fields.

We now extend the dynamical resummation method implemented in ref.\cite{infrared}, and based on the treatement in references\cite{boyww,herringfer}   to the cosmological setting.  As discussed in these references, this method is manifestly unitary and leads to a non-perturbative systematic description of transition amplitudes and probabilities directly in real time,   as shown in ref.\cite{infrared} it is equivalent to the dynamical renormalization group. Here we describe the main aspects of its implementation within the cosmological setting.

 Consider an  interaction picture state $\ket{\Psi_I(\eta)} = \sum_n C_n(\eta)\ket{n}$, expanded in  the Fock states associated with the annihilation and creation operators   of the free field expansions (\ref{fige},\ref{chief}) for each field. To leading order in the adiabatic approximation, these are instantaneous eigenstates of $H_{0}(\eta)$.  Inserting this expansion  into \eqref{Uip} yields an \emph{exact}  set of coupled  equations for the coefficients
\begin{equation}
i\frac{d}{d\eta} C_n(\eta) = \sum_m C_m(\eta)\bra{n}H_I(\eta)\ket{m}.
\end{equation}

\begin{figure}[ht!]
\begin{center}
\includegraphics[height=3.5in,width=3.5in,keepaspectratio=true]{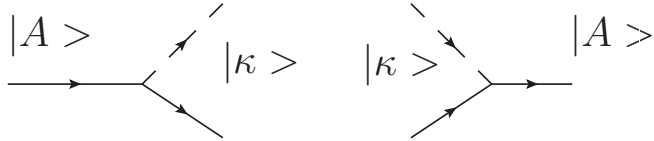}
\caption{Transitions $|A\rangle \leftrightarrow |\kappa\rangle$ in first order in $H_I$.}
\label{fig1:coupling}
\end{center}
\end{figure}

In principle this is an infinite hierarchy of integro-differential equations for the coefficients $C_n(\eta)$; progress is made by truncating the hierarchy to states connected
by the interaction Hamiltonian to a given order in the interaction. Consider that at an initial (conformal) time $\eta_i$  the state is $\ket{A}$ so that $C_A(\eta_i) =C^{(i)}_A$ and $C_{\kappa}(\eta_i) =0$ for $\ket{\kappa}\neq \ket{A}$, and consider a  first order transition  process
$\ket{A}\rightarrow\ket{\kappa}$  to    intermediate multiparticle states $\ket{\kappa}$ with transition matrix elements $\langle \kappa|H_I(\eta)|A\rangle$. Obviously the state $\ket{\kappa}$ will be connected via $H_I(\eta)$ to other multiparticle states $\ket{\kappa'}$  different from $\ket{A}$. Hence for example up to second order in the interaction, the  state $\ket{A}\leftrightarrow  \ket{\kappa}\leftrightarrow \ket{\kappa'}$. Restricting the hierarchy to \emph{first order transitions} from the initial state $\ket{A} \leftrightarrow \ket{\kappa}$ results in the following set of coupled  equations
\bea
i\frac{d}{d\eta} C_A(\eta) &= &  \sum_\kappa C_\kappa(\eta)\bra{A}H_I(\eta)\ket{\kappa}~~;~~C_A(\eta_i)\equiv C^{(i)}_A \label{eqA}\\
i\frac{d}{d\eta} C_\kappa(\eta) &= & C_A(\eta)\bra{\kappa}H_I(\eta)\ket{A}~~;~~ \,C_{\kappa}(\eta_i) =0 \,. \label{eqK}
\eea These processes are shown in fig. (\ref{fig1:coupling}). The initial condition in eqn. (\ref{eqA}) allows for an arbitrary initial amplitude of the state $\ket{A}$, the origin of the initial amplitude will be discussed below (see discussion after eqn. (\ref{entanbos})).

\begin{figure}[ht!]
\begin{center}
\includegraphics[height=3.5in,width=3.5in,keepaspectratio=true]{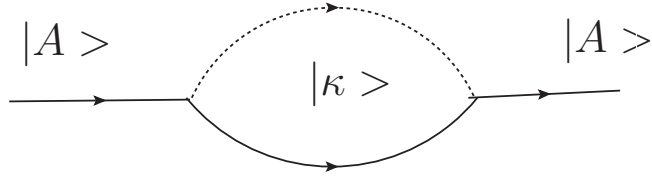}
\caption{One loop self energy corresponding to the state $\ket{A}$.}
\label{fig2:selfenergy}
\end{center}
\end{figure}

 Equation (\ref{eqK}) with $C_{\kappa}(\eta_i)=0$ is formally solved by
 \be C_{\kappa}(\eta) = -i\int^{\eta}_{\eta_i}\,\bra{\kappa}H_I(\eta')\ket{A}\,C_{A}(\eta')\,d\eta' \,, \label{kappapop}\ee
 and inserting this  solution into equation (\ref{eqA}) we find
\begin{equation}\label{diffeqCA}
\frac{d}{d\eta} C_A(\eta) = -\int_{\eta_i}^\eta d\eta' \,
\Sigma_A(\eta,\eta')~ C_A(\eta')\,,
\end{equation} where we have introduced   the \emph{self-energy}
\be \Sigma_A(\eta;\eta') =
\sum_\kappa \bra{A}H_I(\eta)\ket{\kappa}
\bra{\kappa}H_I(\eta')\ket{A} \,. \label{sigmaA}\ee shown in fig.(\ref{fig2:selfenergy}).
 This integro-differential equation  with \emph{memory} yields a non-perturbative solution for the time evolution of the amplitudes and probabilities. In Minkowski space-time and in frequency space, this is recognized as a Dyson resummation of self-energy diagrams, which upon Fourier transforming back to real time, yields the usual exponential decay law\cite{boyww}. Introducing the solution for $C_A(\eta)$ back into (\ref{eqK}) yields the amplitude of the state  $\ket{\kappa}$.

The equation (\ref{diffeqCA}) is in general very difficult to solve exactly, but  a weak coupling assumption yields to a systematic  approximation, achieved by introducing
\be
\mathcal{E}_A(\eta,\eta')  \equiv \int_{\eta_i}^{\eta'} \Sigma_A(\eta,\eta'')\,d\eta'' \,, \label{Evar}
\ee
such that
\be \frac{d}{d\eta'}\,\mathcal{E}_A(\eta,\eta')=\Sigma_A(\eta,\eta') \,,  \label{trick}\ee with the condition
\be   \mathcal{E}_A(\eta,\eta_i)=0 \,. \label{condiEA}\ee

Then \eqref{diffeqCA} can be written as
\begin{equation}
\frac{d}{d\eta} C_A(\eta)  =  -\int_{\eta_i}^\eta d\eta' \,
\frac{d}{d\eta'}\mathcal{E}_A(\eta,\eta')\,  C_A(\eta')
\end{equation}
which can be integrated by parts to yield
\begin{equation}
\frac{d}{d\eta} C_A(\eta) = -\mathcal{E}_A(\eta,\eta) C_A(\eta) +\int_{\eta_i}^\eta d\eta' \,
\mathcal{E}_A(\eta,\eta') \frac{d}{d\eta'}C_A(\eta').
\end{equation}
Since $\mathcal{E}_A \propto \mathcal{O}(H^2_I)$ the first term on the right hand side is of order $H^2_I$, whereas the second is $\order{H^4_I}$ because $dC_A(\eta)/d\eta  \propto \mathcal{O}(H^2_I)$. Therefore  to leading order in the interaction ($\mathcal{O}(H^2_I)$), the evolution equation for the amplitude becomes
\begin{equation}
\frac{d}{d\eta} C_A(\eta) = -\mathcal{E}_A(\eta,\eta) C_A(\eta)  \,,\label{LOeq}
\end{equation} with   solution
\begin{equation}
C_A(\eta) = \exp\Bigl(-\int^\eta_{\eta_i}\mathcal{E}_A(\eta',\eta')\,d\eta'\Bigr)~C^{(i)}_A\,. \label{CAfina}
\end{equation}

This expression  highlights  the non-perturbative nature of the dynamical resummation method. The imaginary part of the self energy $\Sigma_A$ yields a \emph{renormalization} of the frequencies which we will not pursue here\cite{boyww,herringfer}, whereas the real part gives the decay rate, with
\be |C_A(\eta)|^2 = e^{-\int^\eta_{\eta_i} \Gamma_A(\eta') d\eta'} ~|C^{(i)}_A|^2  ~~;~~ \Gamma_A(\eta) = 2 \int^\eta_{\eta_i}d\eta_1\,\mathrm{Re}\,[\Sigma_A(\eta,\eta_1)]   \,. \label{probbaA}\ee

Finally, the time evolution of the amplitude of the  state $\ket{\kappa}$ is obtained by inserting the amplitude (\ref{CAfina})  into (\ref{kappapop}), yielding
\be C_{\kappa}(\eta) =  -i\, C^{(i)}_A \,\int^{\eta}_{\eta_i}\,\bra{\kappa}H_I(\eta')\ket{A}\,
\exp\Bigl(-\int^{\eta'}_{\eta_i}\mathcal{E}_A(\eta'',\eta'')\,d\eta''\Bigr)\,d\eta'  \,. \label{ckapfini}\ee

The hermiticity of $H_I$ leads to the result
\be \frac{d}{d\eta}\,\Bigg\{|C_A(\eta)|^2 + \sum_\kappa |C_\kappa(\eta)|^2 \Bigg\} = 0  \Rightarrow |C_A(\eta)|^2 + \sum_\kappa |C_\kappa(\eta)|^2=|C^{(i)}_A |^2\,, \label{totder}\ee where we used the initial conditions $C_A(\eta_i)=C^{(i)}_A~;~C_\kappa(\eta_i)=0$. This is the statement of unitarity: in  the interaction picture the time evolved state is given by
\be  \ket{\Psi_I(\eta)}   =  U_I(\eta,\eta_i)\ket{\Psi_I(\eta_i)} =  C_A(\eta)\ket{A}+ \sum_{\kappa} C_{\kappa}(\eta)\,\ket{\kappa}  \,,\label{stateIeta}\ee therefore,
\bea \langle \Psi_I(\eta)|\Psi_I(\eta) \rangle & =  & \langle \Psi_I(\eta_i)|U^{\dagger}_I(\eta,\eta_i)\,U(\eta,\eta_i)|\Psi_I(\eta_i) \rangle =  \big|C^2_A(\eta)\big|^2 + \sum_{\kappa} \big|C^2_{\kappa}(\eta)\big|^2 \nonumber \\ & = & \langle \Psi_I(\eta_i)|\Psi_I(\eta_i) \rangle = \big|C^{(i)}_A\big|^2 \,.\label{unitarii} \eea

In our study, for the bosonic case the state $\ket{A}=a^\dagger_{\vk}\,\ket{0^{\varphi};0^{\chi}}\equiv \ket{1^{\varphi}_{\vk};0^{\chi}}$  and the intermediate state $\ket{\kappa} = a^\dagger_{\vp}\,c^\dagger_{\vq} \ket{0^{\varphi};0^{\chi}}\equiv \ket{1^{\varphi}_{\vp};1^{\chi}_{\vq}}$,
 therefore we identify $|C_{\kappa}(\eta)|^2$ as the production probability of the massless particle. This interpretation will be confirmed by the analysis of the expectation value of the energy momentum tensor in this time evolved state in section (\ref{subsec:emts}).  We notice that the production probability of the massless particle is proportional to $|C^{(i)}_A |^2$ (see eqn. (\ref{ckapfini})) which can be associated with the initial ``population'' of the single massive particle state, however, we show in section (\ref{subsec:emts}) that the expectation value of the energy momentum tensor does not depend on this initial condition.

We first describe the dynamical resummation method for the bosonic case, adapting it to the fermionic case in section (\ref{sec:fermions}). For the bosonic model (\ref{conformalaction}), the matrix elements that enter in the self-energy (\ref{sigmaA}) are given    by
\be \bra{1^{\varphi}_{\vp};1^{\chi}_{\vq}}H_I(\eta')\ket{1^{\varphi}_{\vk}} = \frac{\lambda\,a(\eta')}{V^{1/2}}\,  {g_k(\eta')\, g^*_{p}(\eta')\,f_{\vq}(\eta')}\,\delta_{\vk,\vp+\vq}\,, \label{matx1}\ee
 \be \bra{1^{\varphi}_{\vk}}H_I(\eta)\ket{1^{\varphi}_{\vp};1^{\chi}_{\vq}} = \frac{\lambda\,a(\eta)}{V^{1/2}}\, {g^*_k(\eta')\, g_{p}(\eta')\,f^*_{\vq}(\eta')} \,\delta_{\vk,\vp+\vq}\,, \label{matx2}\ee with
  \be \Sigma_k(\eta,\eta') = \sum_{\vp} \bra{1^{\varphi}_{\vk}}H_I(\eta)\ket{1^{\varphi}_{\vp};1^{\chi}_{\vq}}
 \bra{1^{\varphi}_{\vp};1^{\chi}_{\vq}}H_I(\eta')\ket{1^{\varphi}_{\vk}}\,.\label{sigk1}\ee

 In these expressions we have displayed the general form of the matrix elements in terms of the mode functions exact solutions of the free field equations of motion (\ref{equvarfi},\ref{equchi}) to highlight the complexities of the self-energies in curved space time. Obviously the calculation of the self-energy in the general case with the exact solutions of the mode equations is a daunting task, instead we rely on     the adiabatic approximation.

 To leading (zeroth) order in the adiabatic approximation with $g_k(\eta)$ given by (\ref{zerog}) and $f_k(\eta)$ by (\ref{fofk}), summing over the intermediate states and taking the infinite volume limit, we find
\be \Sigma_{k}(\eta,\eta')=\frac{\lambda}{8}\, a(\eta) a(\eta') \int \frac{d^3p}{(2\pi)^3}\,
\frac{e^{i \int^{\eta}_{\eta'} \big[\Omega_k(\eta'')-\Omega_p(\eta'')-q \big]\,d\eta''}}{q\Big[\Omega_k(\eta)\Omega_k(\eta')\Omega_p(\eta)\Omega_p(\eta')\Big]^{1/2}}~~;~~ \vq = \vk-\vp \,,
\label{sigkfin} \ee and  the rate of decay of the initial probability is given by the time integral (\ref{probbaA}). While the conformal time integral of the frequencies can be obtained in closed form\cite{decaycosmo}, neither the momentum integral nor the final time integral leading to the rate $\Gamma(\eta)$ can be done in closed form. A numerical study is not feasible either because of the enormous range in momenta and time. Instead we will leverage the adiabatic approximation to obtain $\Gamma(\eta)$.

The analysis begins by establishing that the self-energy kernel $\Sigma_k(\eta,\eta')$ is short-ranged in the sense that it is dominated by the region $\eta\simeq \eta'$. To see this clearly, let us write
\be \Sigma_k(\eta,\eta') = \frac{\lambda}{8}\, a(\eta) a(\eta') \frac{e^{i \int^{\eta}_{\eta'} \Omega_k(\eta'')\,d\eta''}}{\Big[\Omega_k(\eta)\Omega_k(\eta')\Big]^{1/2}}\, I_k(\eta,\eta') \,,
\label{kerI}\ee  with
\be I_k(\eta,\eta') = \int \frac{d^3p}{(2\pi)^3}\,
\frac{e^{-i \int^{\eta}_{\eta'} \big[\Omega_p(\eta'')+|\,\vk-\vp\,| \big]\,d\eta''}}{|\vk-\vp|\,\Big[
\Omega_p(\eta)\Omega_p(\eta')\Big]^{1/2}}\,. \label{Iker}\ee

Consider first the equal time limit $\eta =\eta'$ for which
\be I_k(\eta,\eta)= \int \frac{d^3p}{(2\pi)^3}\,
\frac{1}{|\vk-\vp|\,\Omega_p(\eta)}   \,,
\label{Iketa} \ee is ultraviolet linearly divergent. The  kernel $I_k(\eta,\eta')$  in (\ref{kerI}) can be calculated explicitly for $M=0$ (see ref.\cite{herringfer})  in which case one finds
 \be I_k(\eta,\eta') \propto \frac{1}{\eta-\eta'}\,, \label{IzeroM} \ee whose divergence as $\eta \rightarrow \eta'$ reflects the linear ultraviolet divergence. This short time divergence is
 independent of the mass, therefore the full kernel $I_k(\eta,\eta')$ for $M\neq 0$ is expected to feature this short time behavior. Motivated by this observation we seek an expansion anchored in the adiabatic approximation, this is achieved by writing
 \be \Omega_p(\eta') = \Bigg[ p^2 + M^2 a^2(\eta) + M^2 a^2(\eta)\,\Big[ \Big(\frac{ \eta-\eta' }{\eta}\Big)^2 - 2 \,\Big(\frac{ \eta-\eta' }{\eta}\Big)\Big]   \Bigg]^{1/2} \,, \label{omex}\ee introducing $\tau = \Omega_k(\eta)(\eta-\eta')$ it follows that (\ref{omex}) becomes
 \be \Omega_p(\eta') = \Omega_p(\eta) \Bigg[1 + \frac{1}{\gamma^2_p(\eta)} \Big( -2\,\epsilon_k(\eta) \tau + \epsilon^2_k(\eta) \tau^2 \Big)   \Bigg]^{1/2} \simeq \Omega_p(\eta) \Bigg[1 - \frac{\epsilon_k(\eta)}{\gamma^2_p(\eta)} \,  \tau + \cdots \Bigg] \,,\label{omex2}\ee
 where $\gamma_p(\eta) = \Omega_p(\eta)/M a(\eta)$ is the local Lorentz factor,   $\epsilon_k(\eta) = 1/(\Omega_k(\eta)\eta)\ll 1$ is the dimensionless adiabatic parameter introduced in eqn. (\ref{epsi}), and only  displayed the first order term in the expansion in $\epsilon_k(\eta)$ in (\ref{omex2}). We confirm below self-consistently that for $\tau \simeq 1/\epsilon_k(\eta)$ when the higher order adiabatic terms in (\ref{omex2}) become of the same order as the leading contribution, the kernel (\ref{Iker}) is suppressed by $\propto \epsilon^2_k(\eta)$, therefore confirming the consistency of  the leading order terms in this expansion. We now proceed to prove this important aspect self-consistently.

Up to first order in $\epsilon_k(\eta)$ (\ref{Iker}) becomes

\be I_k(\eta,\eta') = \int \frac{d^3p}{(2\pi)^3}\,\Big[1+ \frac{\epsilon_k(\eta)\,\tau}{2\gamma^2_p(\eta)} \Big]\,
\frac{e^{-i \Big\{ \big[\Omega_p(\eta)+|\vp-\vk| \big](\eta-\eta')\big[1 - \delta_{p,k}(\eta)\,\epsilon_k(\eta)\,\tau \big]\Big\}}}{|\vp-\vk|\,
\Omega_p(\eta) }\,, \label{Iker1or}\ee where
\be \delta_{p,k}(\eta) = \frac{\Omega_p(\eta)}{2\gamma^2_p(\eta)(\Omega_p(\eta)+|\vp-\vk|)} < \frac{1}{2} \,.\label{delpk}\ee Obviously even at this first order in $\epsilon_k(\eta)$ the integral cannot be done in closed form, however, it allows us to understand the range of the kernel. First, since $\delta_{p,k}(\eta) < 1/2$ at all times and for all values of $p$ we approximate it as $\delta_{p,k}(\eta)\equiv \overline{\delta} < 1/2$ for all momenta and time, similarly, we replace $\gamma_p(\eta) \equiv \overline{\gamma} \geq 1$ for all values of momenta and time, and finally we introduce $T = (\eta-\eta')\Big[1- \overline{\delta}\,\epsilon_k(\eta)\,\tau \Big]$. With these approximations,
\be I_k(\eta,\eta') \equiv \Big[1+ \frac{\epsilon_k(\eta)\,\tau}{2\overline{\gamma}^2} \Big]\, \overline{I}_k(\eta,\eta') ~~;~~ \overline{I}_k(\eta,\eta')=\int \frac{d^3p}{(2\pi)^3}\,
\frac{e^{-i  \big[\Omega_p(\eta)+|\vp-\vk| \big]T}}{|\vp-\vk|\,\,
\Omega_p(\eta) }\,, \label{Ikerap}\ee  Introducing the spectral density
\be \rho(k_0;k) = \int \frac{d^3p}{(2\pi)^3}\,
\frac{\delta(k_0- \Omega_p(\eta)-|\vp-\vk|)}{|\vp-\vk|\,
\Omega_p(\eta) }\,, \label{rho}\ee which depends on $\eta$ parametrically, we can write
\be \overline{I}_k(\eta,\eta')= \int^\infty_{-\infty} \rho(k_0;k) e^{-ik_0 T} \, dk_0\,. \label{specrep}\ee The spectral density (\ref{rho}) is the same as that  found
in the study of infrared dynamics in Minkowski space time in ref.\cite{infrared}, but depending parametrically on $\eta$,  it is given by
\be \rho(k_0;k) = \overline{\rho}(k_0;k)\, \Theta(k_0-\Omega_k(\eta))~~;~~ \overline{\rho}(k_0;k)=\frac{1}{4\pi^2}  \Bigg[\frac{k^2_0-\Omega^2_k(\eta)}{k^2_0 - k^2} \Bigg]\,.  \label{rhomink}\ee The $T\rightarrow 0$ limit of (\ref{specrep}) is determined by the large $k_0$ behavior of the spectral density\footnote{This can be seen by rescaling $k_0 T = \zeta$ in the integral  in (\ref{specrep}).}, introducing a convergence factor $T \rightarrow T-i\varepsilon, \varepsilon \rightarrow 0^+$, we find
\be \overline{I}_k(T \rightarrow 0) = \frac{-i}{4\pi^2\,T}\,e^{-i\Omega_k(\eta) T} \,,\label{shorti}\ee  which reflects the short time behavior (\ref{IzeroM}). The asymptotic long time limit $T \rightarrow \infty$ can be obtained systematically as follows:   using the identity
\be e^{-ik_0 T} = \frac{i}{T} \frac{d}{dk_0} \Big( e^{-ik_0 T} \Big)\,\label{iden}  \ee   integrate by parts (with the convergence factor). Because the spectral density vanishes at   threshold $k_0 = \Omega_p(\eta)$   this procedure must be repeated for a second time obtaining
\be \overline{I}(T\rightarrow \infty) = \frac{e^{-i\Omega_k(\eta) T} }{T^2} \, \frac{d\overline{\rho}}{dk_0}\Big|_{k_0=\Omega_k(\eta)}+ \mathcal{O}(1/T^3)\,.  \label{largeT}\ee  This result is important: in terms of   $\tau=\Omega_k(\eta)(\eta-\eta')$   it follows that

\be \overline{I}(T\rightarrow \infty) \propto \frac{\Big(\epsilon_k(\eta)\,\Omega_k(\eta)\Big)^2}{\Big(\epsilon_k(\eta)\,\tau\Big)^2}\, \frac{1}{\Big[1- \overline{\delta}\,\epsilon_k(\eta)\,\tau \Big]^2} \ee therefore, for $\epsilon_k\,\tau \simeq 1 $ when the higher adiabatic orders become important, the kernel $I_k(\eta,\eta') \propto \epsilon^2_k(\eta)$, hence of subleading adiabatic order.

The main conclusion of this analysis is that the self-energy kernel is short
ranged in time, and to leading adiabatic order it is the region $\eta \simeq \eta'$ that is dominant. At the time scale when the higher order adiabatic terms become comparable to the zeroth order the kernel is suppressed by a high power of $\epsilon$. Therefore terms with powers of  $\epsilon_k(\eta)\, \tau $ can be safely neglected to leading adiabatic order, thereby validating keeping the zeroth adiabatic order in the analysis below.

  Armed with this   result, we can now focus on the leading contribution to the self-energy $\Sigma_k(\eta,\eta')$ in (\ref{kerI}). To leading   order the expansion (\ref{omex2}) yields  $\Omega_k(\eta') =  \Omega_k(\eta) + \cdots$, furthermore, using the identity (valid during RD)
\be a(\eta') = a(\eta)\Big[ 1- \Big( \frac{\eta-\eta'}{\eta}\Big)\Big]  = a(\eta)\Big[ 1- \epsilon_k(\eta) \,\tau\Big] = a(\eta) + \cdots \label{aiden}\ee  and from eqn. (\ref{omex2}) $\Omega_k(\eta') = \Omega_k(\eta) + \cdots $ where the $\cdots$ stand for higher order terms in the adiabatic expansion, we finally find, to leading adiabatic order
\be \Sigma_k(\eta,\eta') = \frac{\lambda^2\,a^2(\eta)}{8\,\Omega_k(\eta)} \,\int^\infty_{\Omega_k(\eta)}  \overline{\rho}(k_0;k)\, e^{-i(k_0-\Omega_k(\eta))(\eta-\eta')}\,dk_0 \,. \label{sigkerzero}\ee  where $\overline{\rho}(k_0;k)$ is given by eqn. (\ref{rhomink}). The integral in $\eta'$ can now be carried out.

The decay rate of the single $\varphi$ particle of comoving momentum $k$, given by eqn. (\ref{probbaA}) is
\be \Gamma_{k}(\eta)= \frac{\lambda^2\,a^2(\eta)}{4\,\Omega_k(\eta)}\, \int^\infty_{\Omega_k(\eta)}\,\overline{\rho}(k_0;k)\, \frac{\sin\Big[(k_0-\Omega_k(\eta))(\eta-\eta_i) \Big]}{k_0-\Omega_k(\eta)}\,dk_0\,.  \label{ratafi}\ee

Introducing the dimensionless variable
\be s = \frac{k_0-\Omega_k(\eta)}{\Omega_k(\eta)}\,, \label{svar} \ee which depends explicitly on $\eta$ (we suppressed the argument), it follows that the spectral density (\ref{rhomink}) written in terms of $s$, vanishes linearly in $s$ and restoring its dependence on $\eta$ can be written as
\be \overline{\rho}(s;\eta) = s \,\mathcal{D}(\eta)\,\Big( 1 + s \,\widetilde{\sigma}(s;\eta)\Big)~~;~~ \mathcal{D}(\eta) = \frac{d\,\overline{\rho}(s;\eta)}{ds}\Big|_{s=0}\,, \label{tilsigdef}\ee where $\widetilde{\sigma}(0;\eta)$ is time dependent but finite. For $\overline{\rho}(k_0,k)$ given by (\ref{rhomink}) we find
\be \mathcal{D}(\eta) = \frac{\gamma^2_k(\eta)}{2\pi^2}~~;~~ \widetilde{\sigma}(s;\eta) = \frac{1}{2}\, \Bigg[\frac{\frac{1}{\gamma^2_k(\eta)} -4 -2s}{\frac{1}{\gamma^2_k(\eta)} +2s+s^2 }\Bigg]\,. \label{parabos}\ee

The rate (\ref{ratafi}) can now be written as
\be \Gamma_{k}(\eta)= \frac{\lambda^2\,a^2(\eta)\,\mathcal{D}(\eta)}{4\,\Omega_k(\eta)}\, \int^{\infty}_{0} \,\Big( 1 + s \,\widetilde{\sigma}(s;\eta)\Big) \, \sin\big[s \,\Omega_k(\eta)(\eta-\eta_i)\big]\,ds \,. \label{ratafis}\ee

In Minkowski space-time the region $s\simeq 0$ yields an infrared divergence in the long time limit\cite{infrared}, this is also the case in the (RD) cosmology as is made explicit by the following analysis. Let us  write: $\int^\infty_0 (\cdots) ds = \int^1_0 (\cdots) ds + \int^\infty_1 (\cdots)ds$,  yielding
\be \Gamma_k(\eta) = \Gamma^{(1)}_k(\eta)+
 \Gamma^{(2)}_k(\eta)+\Gamma^{(3)}_k(\eta)\,,\label{gamas} \ee with
 \bea \Gamma^{(1)}_k(\eta) & = & \frac{\lambda^2\,a^2(\eta)\,\mathcal{D}(\eta)}{4\,\Omega_k(\eta)}\,
\int^1_0 \,\sin\big[s \,\Omega_k(\eta)(\eta-\eta_i)\big]\,ds \, \,, \label{gama1} \\
\Gamma^{(2)}_k(\eta) & = & \frac{\lambda^2\,a^2(\eta)\,\mathcal{D}(\eta)}{4\,\Omega_k(\eta)}\,
\int^1_0\, s\, \widetilde{\sigma}(s;\eta)\,\sin\big[s \,\Omega_k(\eta)(\eta-\eta_i)\big]\,ds \,, \label{gama2} \\
\Gamma^{(3)}_k(\eta) & = & \frac{\lambda^2\,a^2(\eta)\,}{4\,\Omega_k(\eta)}\,
\int^\infty_1\, \frac{\overline{\rho}(s;\eta)}{s}\,\sin\big[s \,\Omega_k(\eta)(\eta-\eta_i)\big]\,ds \,,\label{gama3} \eea obviously the first integral (\ref{gama1}) is straightforward.
Finally, from eqn. (\ref{probbaA}) to understand the time evolution of the survival probability of the initial state, we need the $\eta$-integral $\int^\eta_{\eta_i}\,\Gamma_k(\eta')\,d\eta$.  The contribution from $\Gamma^{(1)}_k(\eta)$ is shown below to be infrared divergent in the long time limit, whereas those from $\Gamma^{(2,3)}_k(\eta)$ are infrared and ultraviolet finite and feature a slow time evolution in the long time limit. Their contribution is analyzed in detail in appendix (\ref{appendix:gamma23}).

Carrying out the $s$- integration for the first contribution, we find
\be \int^\eta_{\eta_i}\,\Gamma^{(1)}_k(\eta')\,d\eta' = 2\,\Delta_b \, \int^{\eta}_{\eta_i}\,\frac{\Big[1-\cos\Big(\Omega_k(\eta')\,(\eta'-\eta_i) \Big) \Big]}{(\eta'-\eta_i)}\,d\eta'\,,\label{cont1} \ee where we introduced the effective dimensionless coupling
\be \Delta_b = \Big(\frac{\lambda}{4\pi\,M}\Big)^2\,.  \label{Deltadef}\ee This integral cannot be done in closed form, however, it can be obtained in an adiabatic expansion as follows: with the definition
\be x \equiv \Omega_k(\eta')\,(\eta'-\eta_i) \Rightarrow \frac{dx}{d\eta'} = \Omega_k(\eta')\Big[1+\widetilde{\epsilon}_k(\eta')\,x\Big]\,, \label{xdefs}\ee where
 $\widetilde{\epsilon}_k$ is given by eqn. (\ref{epsi}) in terms of the adiabatic ratio $\epsilon_k$. In the above expressions $\eta'$ is a function of $x$. In terms of this variable and taking $\eta \gg \eta_i$ we find
 \be \int^\eta_{\eta_i}\,\Gamma^{(1)}_k(\eta')\,d\eta' = 2\,\Delta_b \, \Bigg[
 \underbrace{\int^{1/\epsilon_k(\eta)}_{0}\,\frac{1-\cos(x)}{x} \,dx }_{A} - \underbrace{\int^{1/\epsilon_k(\eta)}_{0}\,\widetilde{\epsilon_k}(x)\,\frac{1-\cos(x)}{1+\widetilde{\epsilon}_k(x)\,x} \,dx }_B \Bigg]\,, \label{finint} \ee
 the (A) integral in the long time limit $1/\epsilon_k(\eta) = \Omega_k(\eta)\, \eta \rightarrow \infty$ becomes
 \be (A) \rightarrow \ln\Big[\Omega_k(\eta)\eta\Big]\,, \label{Aint} \ee whereas for the (B) term, the cosine term averages out, furthermore  note that at $x = 1/\epsilon_k(\eta) $ the ratio $\widetilde{\epsilon}_k(\eta)/\epsilon_k(\eta) = 1/\gamma^2_k(\eta) \leq 1$, therefore $(B)\simeq \mathcal{O}(1)$ and varies slowly in the long time limit (keeping $\widetilde{\epsilon} \simeq \mathrm{constant}$ it follows that $(B) \leq \ln(2)$). In appendix (\ref{appendix:gamma23}) we show that the contributions from $\Gamma^{(2,3)}_k$ are infrared and ultraviolet finite and remain bound and slowly varying in time, reaching a constant value at asymptotically long time.
 Therefore, we find for $\Omega_k(\eta)\,  \eta = E_k(t)/H(t) \gg 1$
 \be   \int^\eta_{\eta_i}\,\Gamma^{}_k(\eta')\,d\eta' = 2\,\Delta_b \,\ln\Big[ \frac{E_k(t)}{H(t)}\Big]+ z(t)\,, \label{asyintgama}\ee where $z(t)$ is a slowly varying function of time that approaches an infrared and ultraviolet finite constant in the asymptotic long time limit (see appendix (\ref{appendix:gamma23})).

 During the (RD) era $H(t) = 1/2\,t$, therefore in terms of cosmic time, the contribution that grows in time on the right hand side of (\ref{asyintgama}) is $ 2\Delta_b \,\ln[2\,E_k(t)t]$ which is very similar to the result in Minkowski space time\cite{infrared}, however, in the expanding cosmology the local energy depends on time as a consequence of the cosmological redshift. With the scale factor given by (\ref{etaoft}) it is convenient to introduce the time scale $t_{nr}$ that determines when the particle becomes non-relativistic as
 \be t_{nr} = \frac{k^2}{2M^2H_R}\,, \label{tnr} \ee so that the local Lorentz factor
 \be \gamma_k(t) = \sqrt{1+\frac{t_{nr}}{t}} \Rightarrow \Bigg\{
  \begin{array}{c}
   \mathrm{Relativistic}~~\mathrm{for}~~t\ll t_{nr} \\
   \mathrm{Non-relativistic} ~~\mathrm{for}~~ t \gg t_{nr}
                                                                 \end{array}
 \,.\label{gamtnr}\ee Hence,   we find asymptotically
 \be   \int^\eta_{\eta_i}\,\Gamma_k(\eta')\,d\eta' = 2\,\Delta_b \,\ln\Big[2M\,t \gamma_k(t)\Big]  + z(t)   \,. \label{asyx} \ee

 In summary, the survival probability of a single $\varphi$ particle state with momentum $\vk$ in the long time limit is
 \be |C^{\varphi}_k(t)|^2 =  C^{\varphi}_k(t_i)\, \Big[ \frac{E_k(t)}{H(t)}\Big]^{-2\Delta_b}\,\mathcal{Z}(t) ~~;~~ \mathcal{Z}(t) = e^{-z(t)}\,, \label{surviprob}\ee or in terms of cosmic time,
 that
 \be  |C^{\varphi}_k(t)|^2 \simeq C^{\varphi}_k(t_i)\, \Big[2M\,t\,\gamma_k(t)\Big]^{-2\Delta_b}\,\mathcal{Z}(t)\,. \label{almink}\ee The wave function renormalization $\mathcal{Z}(t)$ is a slowly varying function of time that remains bounded at long time. The cosmological redshift responsible for the time dependence of the local Lorentz factor entails a crossover of the decaying term:
 \be |C^{\varphi}_k(t)|^2 \propto \Bigg\{ \begin{array}{c}
                                            t^{-\Delta_b} ~~;~~ t \ll t_{nr} \\
                                           t^{-2\Delta_b} ~~;~~ t \gg t_{nr}
                                          \end{array}
\,. \label{crosstime}
 \ee

The anomalous dimension $2\Delta_b$ is the same as in Minkowski space-time  and originates in the infrared divergence\cite{infrared}.

The amplitude of the multiparticle state $\ket{1^{\varphi}_{\vp};1^{\chi}_{\vk-\vp}}$ is
\be C^{\varphi,\chi}_{\vp;\vk}(\eta)= -i \frac{\lambda}{V^{1/2}}\,\int^{\eta}_{\eta_i} a(\eta')\, \frac{e^{-i\int^{\eta'}_{\eta_i}\big(\Omega_k(\eta'')-\Omega_p(\eta'')\big)\,d\eta'' }\,e^{i|\vk-\vp|\eta'}}{\Big[8\,\Omega_k(\eta')\Omega_p(\eta')|\vk-\vp|\Big]^{1/2}}\,\,C^{\varphi}_k(\eta') \, d\eta' \,, \label{Cmultibos}\ee and the time evolved state in the interaction picture
is given by
\be
\ket{\Psi_I(\eta)} = C^{\varphi}_k(\eta)\,\ket{1^{\varphi}_{\vk};0^{\chi}}+ \sum_{\vp} C^{\varphi,\chi}_{\vp,\vk}(\eta)\,\ket{1^{\varphi}_{\vp};1^{\chi}_{\vk-\vp}}\,
\label{entanbos}\ee unitarity (\ref{totder}) implies that $  \langle \Psi_I(\eta)|\Psi_I(\eta)\rangle  = |C^{\varphi}_k(\eta_i)|^2$.

The second term in (\ref{entanbos}) describes an entangled state of the single $\varphi$ particle and a single $\chi$ particle, this cloud of $\chi$ particles ``dresses'' the $\varphi$ particle. Since $ C^{\varphi}_k(\eta) \rightarrow 0$ as $\eta \rightarrow \infty$ only the second term survives in the asymptotic long time limit, hence the sum rule (\ref{unitarii}) yields $\sum_{\vp} |C^{\varphi,\chi}_{\vp,\vk}(\eta)|^2 = |C^{\varphi}_k(\eta_i)|^2$ thus saturating the unitarity constraint in the asymptotic long time limit.

The initial amplitude $C^{\varphi}_k(t_i)$ must be determined from the amplitude of the single particle state at the   time when the adiabatic approximation begins to be valid. It is determined by the processes that lead to the production of single $\varphi$ particle states prior to the onset of the adiabatic era, such as   particle production
during inflation or the post-inflationary era. However, we show in section (\ref{sec:emt}) that the expectation value of the energy momentum tensor does not depend on this initial condition.

\section{Fermionic case:}\label{sec:fermions}

\subsection{Adiabatic approximation for fermions:}

We consider   the  massless scalar field $\pi$  as the ultra light degree of freedom Yukawa coupled to one Dirac fermion in a spatially flat Friedmann-Robertson-Walker (FRW) cosmology.

In comoving
coordinates, and for a (RD) cosmology (with vanishing Ricci scalar) the action is given by
\be
S   =  \int d^3x \; dt \;  \sqrt{|g|} \,\Bigg\{
\frac{1}{2} g^{\mu\nu}\,\partial_\mu \pi \partial_\nu \pi  +
\overline{\Psi}  \Big[i\,\gamma^\mu \;  \mathcal{D}_\mu -M-Y \pi\Big]\Psi     \Bigg\}\,, \label{lagrads}
\ee   Introducing  the vierbein field $e^\mu_a(x)$  defined as
$$
g^{\mu\,\nu}(x) =e^\mu_a (x)\;  e^\nu_b(x) \;  \eta^{a b} \; ,
$$
\noindent where $\eta_{a b}$ is the Minkowski space-time metric,
the curved space time Dirac gamma- matrices $\gamma^\mu(x)$ are given
by

\be
\gamma^\mu(x) = \gamma^a e^\mu_a(x) \quad , \quad
\{\gamma^\mu(x),\gamma^\nu(x)\}=2 \; g^{\mu \nu}(x)  \; ,
\label{gamamtx}\ee where the $\gamma^a$ are the Minkowski space time Dirac matrices, chosen to be in the standard Dirac representation, and the fermionic covariant derivative $\mathcal{D}_\mu$    is given in terms of the spin connection\cite{weinbergbook,birrell,duncan,casta} by

\be
\mathcal{D}_\mu   =    \partial_\mu + \frac{1}{8} \;
[\gamma^c,\gamma^d] \;  e^\nu_c  \; \left(\partial_\mu e_{d \nu} -\Gamma^\lambda_{\mu
\nu} \;  e_{d \lambda} \right) \,,  \label{fermicovader}
\ee  where $\Gamma^\lambda_{\mu
\nu}$ are the usual Christoffel symbols.

With the metric in conformal time given by (\ref{conformalmetric})
  the vierbeins $e^\mu_a$ are given by (up to a local Lorentz transformation)
\be
 e^\mu_a = a^{-1}(\eta)\; \delta^\mu_a ~~;~~ e^a_\mu = a(\eta) \; \delta^a_\mu \,. \label{vierconf}\ee

The fermionic part of the action in conformal coordinates now becomes
\be S_f = \int d^3 x\,d\eta\,a^4(\eta)\,  \overline{\Psi}(\vec{x},\eta)\,\Bigg[ i \frac{\gamma^0}{a(\eta)}\,\Big(\frac{d}{d\eta} + 3 \frac{a^{'}(\eta)}{2a(\eta)}\Big) + i \,\frac{\gamma^i}{a(\eta)}\nabla_i - M - Y\,\pi\Bigg]\Psi(\vec{x},\eta)\,. \label{Sf}\ee

The Dirac Lagrangian density in conformal time and with the conformal rescaling of the $\pi$ field as in eqn. (\ref{rescale}) simplifies to
\be
\sqrt{-g} \; \overline{\Psi}\Big(i \; \gamma^\mu \;  \mathcal{D}_\mu
 -M-Y\phi \Big)\Psi  =
\big(a^{3/2}(\eta)\,\overline{\Psi}(\vec{x},\eta)\big) \;  \Big[i \;
{\not\!{\partial}}-M\,a(\eta)-Y\,\chi)   \Big]
\big(a^{3/2}(\eta)\,{\Psi}(\vec{x},\eta)\big)\,, \label{confferscal}
\ee
\noindent where $i {\not\!{\partial}}=\gamma^a \partial_a$ is the usual Dirac
differential operator in Minkowski space-time in terms of flat
space time $\gamma^a$ matrices. Introducing the conformally rescaled fermionic fields
\be   a^{\frac{3}{2}}(\eta)\,{\Psi(\vx,t)}= \psi(\vx,\eta)\,, \label{rescaledfields}\ee
and neglecting surface terms, the action becomes
   \be  S    =
  \int d^3x \; d\eta \, \Big\{\mathcal{L}_0[\chi]+\mathcal{L}_0[\psi]+\mathcal{L}_I[\chi,\psi] \Big\} \;, \label{rescalagds}\ee
  with
  \bea \mathcal{L}_0[\chi] & = & \frac12\left[
{\chi'}^2-(\nabla \chi)^2  \right] \,, \label{l0chi}\\
\mathcal{L}_0[\psi] & = & \overline{\psi} \;  \Big[i \;
{\not\!{\partial}}- M(\eta)   \Big]
 {\psi}  \,,\label{l0psi}\\ \mathcal{L}_I[\chi,\psi] & = & -Y   \overline{\psi}\,\chi\,\psi \; . \label{lI}\eea    The effective time dependent fermion mass is given by

\be M (\eta) = M \,a(\eta) \,.  \label{masfer}\ee

In the non-interacting case, $Y =0$, the Heisenberg equations of motion for the spatial Fourier modes with comoving wavevector $\vec{k}$ for the conformally rescaled scalar field is given by eqn. (\ref{Eomchi}).

The Heisenberg fields are quantized in a comoving volume $V$,   the real scalar field $\chi$ is  expanded as in eqn. (\ref{chiefmink}), and for Dirac fermions the field $ \psi({\vec x},\eta) $ is expanded  as
\be
\psi (\vec{x},\eta) =    \frac{1}{\sqrt{V}}
\sum_{\vec{k},\lambda=1,2}\,   \left[b_{\vec{k},\lambda}\, U_{\lambda}(\vec{k},\eta)\,e^{i \vec{k}\cdot
\vec{x}}+
d^{\dagger}_{\vec{k},\lambda}\, V_{\lambda}(\vec{k},\eta)\,e^{-i \vec{k}\cdot
\vec{x}}\right] \; ,
\label{psiex}
\ee
where the spinor mode functions $U,V$ obey the  Dirac equations\cite{fercurved1,fercurved2,fercurved3,feradia,feradia2,feradia3,boydVS,baacke,ghosh,landete}
\bea
\Bigg[i \; \gamma^0 \;  \partial_\eta - \vec{\gamma}\cdot \vec{k}
-M(\eta) \Bigg]U_\lambda(\vec{k},\eta) & = & 0 \,,\label{Uspinor} \\
\Bigg[i \; \gamma^0 \;  \partial_\eta + \vec{\gamma}\cdot \vec{k} -M(\eta)
\Bigg]V_\lambda(\vec{k},\eta) & = & 0 \,.\label{Vspinor}
\eea

These   equations become simpler by  writing
\bea
U_\lambda(\vec{k},\eta) & = & \Bigg[i \; \gamma^0 \;  \partial_\eta -
\vec{\gamma}\cdot \vec{k} +M(\eta)
\Bigg]f_k(\eta)\, \mathcal{U}_\lambda \,,\label{Us}\\
V_\lambda(\vec{k},\eta) & = & \Bigg[i \; \gamma^0 \;  \partial_\eta +
\vec{\gamma}\cdot \vec{k} +M( \eta)
\Bigg]h_k(\eta)\,\mathcal{V}_\lambda \,,\label{Vs}
\eea
\noindent with $\mathcal{U}_\lambda;\mathcal{V}_\lambda$ being
constant spinors\cite{boydVS,baacke} obeying
\be
\gamma^0 \; \mathcal{U}_\lambda  =  \mathcal{U}_\lambda
  \qquad , \qquad
\gamma^0 \;  \mathcal{V}_\lambda  =  -\mathcal{V}_\lambda\,. \label{Up}
\ee
Inserting (\ref{Us},\ref{Vs}) into the Dirac equations (\ref{Uspinor},\ref{Vspinor}) and using (\ref{Up}), it follows that the mode functions $f_k(\eta);h_k(\eta)$ obey the
equations
\bea \left[\frac{d^2}{d\eta^2} +
\Omega^2_k(\eta)-i \; M'(\eta)\right]f_k(\eta) & = & 0 \,, \label{modefermionf}\\
\left[\frac{d^2}{d\eta^2} + \Omega^2_k(\eta)+i \; M'(\eta)\right]h_k(\eta)
& = & 0 \,.\label{modefermionh}
\eea
where
\be \Omega_k(\eta) = \sqrt{k^2+M^2(\eta)}\,. \label{ferfreq}\ee

Multiplying the Dirac equations on the left by $\gamma^0$, it is straightforward to confirm that
\be \frac{d}{d\eta} (U^\dagger_\lambda(q,\eta)\,U_\lambda(q,\eta)) =0 ~~;~~ \frac{d}{d\eta} (V^\dagger_\lambda(q,\eta)\,V_\lambda(q,\eta)) =0 \,.\label{constnorm}\ee We choose the normalizations
\be U^\dagger_\lambda(q,\eta)\,U_{\lambda'}(q,\eta) = V^\dagger_\lambda(q,\eta)\,V_{\lambda'}(q,\eta) = \delta_{\lambda,\lambda'}\,, \label{normas}\ee so that the operators $b,b^\dagger, d,d^\dagger$ obey the canonical anticommutation relations. Furthermore, we will choose particle-antiparticle boundary conditions so that $h_k(\eta) = f^*_k(\eta)$. We note that for $M =0$ the conformally rescaled fermi fields obey the same equations as in Minkowski space-time but in terms of conformal time, this is also the case for massless scalar fields in a (RD) cosmology where the Ricci scalar vanishes.  The adiabatic expansion for Fermi fields has been studied in refs.\cite{herringfer,ghosh,landete,fercurved3,feradia,feradia2,feradia3} to which we refer the reader for details. Here we summarize the results up to leading (zeroth) adiabatic order. In particular we
recognize that
\be \frac{M'(\eta)}{\Omega^2_k(\eta)}= \frac{H(t)}{\gamma_k(t)\,E_k(t)}= \frac{\epsilon_k(t)}{\gamma_k(t)}\,, \label{mprimf}\ee therefore the purely imaginary term in the mode equations (\ref{modefermionf},\ref{modefermionh}) are of first adiabatic order and will be neglected to leading (zeroth)  order.

 Hence, to leading order we find
\be f_k(\eta) = h^*_k(\eta) = \frac{e^{-i\,\int^{\eta}_{\eta_i}\,\Omega_k(\eta')\,d\eta'}}{\sqrt{2\,\Omega_k(\eta)}}\,.  \label{wkbfer} \ee  To this order    the Dirac spinor solutions  in the standard Dirac representation and with the normalization conditions (\ref{normas})  are found to be
\bea  U_\lambda(\vk,\eta) & = &   \frac{e^{-i\,\int^{\eta}_{\eta_i}\,\Omega_k(\eta')\,d\eta'}}{\sqrt{2\,\Omega_k(\eta)}}\,\mathcal{U}_\lambda(\vk,\eta)
\label{Uspi}\\
V_\lambda(\vk,\eta) & = & \frac{e^{i\,\int^{\eta}_{\eta_i}\,\Omega_k(\eta')\,d\eta'}}{\sqrt{2\,\Omega_k(\eta)}}\,
\mathcal{V}_\lambda(\vk,\eta) \,\label{Vspi} \eea  where
\be \mathcal{U}_\lambda(\vk,\eta) =  \frac{1}{\sqrt{\mathcal{W}(\eta)}}\, \left(
                            \begin{array}{c}
                               \mathcal{W}(\eta)\,\, \xi_\lambda \\
                               {\vec{\sigma}\cdot \vec{k}} \, \, \xi_\lambda \\
                            \end{array}
                          \right) ~~;~~ \xi_1 = \left(
                                                   \begin{array}{c}
                                                     1 \\
                                                     0 \\
                                                   \end{array}
                                                 \right) \;; \; \xi_2 = \left(
                                                                       \begin{array}{c}
                                                                         0 \\
                                                                         1 \\
                                                                       \end{array}
                                                                     \right) \,,
 \label{Uspinorsol} \ee  and

 \be \mathcal{V}_\lambda(\vk,\eta) =  \frac{1}{\sqrt{\mathcal{W}(\eta)}}\, \left(
                            \begin{array}{c}
                                {\vec{\sigma}\cdot \vec{k}} \, \, \widetilde{\xi}_\lambda \\
                                \mathcal{W}(\eta)\, \,\widetilde{\xi}_\lambda  \\
                            \end{array}
                          \right) ~~;~~ \widetilde{\xi}_1 = \left(
                                                   \begin{array}{c}
                                                    0 \\
                                                     1 \\
                                                   \end{array}
                                                 \right) \;; \; \widetilde{\xi}_2 = -\left(
                                                                       \begin{array}{c}
                                                                         1 \\
                                                                         0 \\
                                                                       \end{array}
                                                                     \right) \,,
 \label{Vspinorsol} \ee
 where we introduced
 \be \mathcal{W}_k(\eta) = \Omega_k(\eta)+M(\eta)= a(\eta)\,[E_k(\eta)+M]\,. \label{Wofeta} \ee

To leading adiabatic order the $\mathcal{U}$  spinors satisfy the completeness relations
\be  \sum_{\lambda=1,2} \mathcal{U}_{\lambda,a}(\vk,\eta)\,\overline{\mathcal{U}}_{\lambda,b}(\vk,\eta')   \equiv (\Lambda^{+}_{k}(\eta,\eta'))_{ab} =    \frac{1}{ \sqrt{ \mathcal{W}_k(\eta) \mathcal{W}_k(\eta')}}\,\,\left(
       \begin{array}{cc}
         \mathcal{W}_k(\eta)\mathcal{W}_k(\eta') \,\mathbb{I} ~ & ~ -\vec{\sigma}\cdot\vk\, \mathcal{W}_k(\eta) \\
         \vec{\sigma}\cdot\vk\, \mathcal{W}_k(\eta') ~ & ~ - {k^2} \,\mathbb{I} \\
       \end{array}
     \right)
  \,,\label{relaspinors}\ee in particular for $\eta = \eta'$
  \be \Lambda^{+}(\eta,\eta)   =
 {\Big[ {\not\!{K}}(\eta) + M(\eta)\Big]} ~~;~~ K_\mu(\eta)= (\Omega_k(\eta),-\vec{k})\,. \label{eqtimrel}
\ee

\subsection{Dynamical resummation:}\label{subsec:drmfer}

We now have the main ingredients to implement the dynamical resummation for this fermionic case, for which  the  single particle initial state is taken to be $\ket{A} = \ket{1^{\psi}_{\vk,\alpha};0^{\chi}}$ and the intermediate state connected to $\ket{A}$ at first order in the interaction is $\ket{\kappa} = \ket{1^{\psi}_{\vp,\beta};1^{\chi}_{\vq}}$. Therefore,  to lowest adiabatic order the transition matrix elements are
\be \bra{1^{\psi}_{\vp,\beta};1^{\chi}_{\vq}}H_I(\eta')\ket{1^{\psi}_{\vk,\alpha}} = \frac{Y}{V^{1/2}}\,\delta_{\vk,\vp+\vq}\, \frac{e^{-i \int^{\eta'}_{\eta_i} \big[\Omega_k(\eta'')-\Omega_p(\eta'') \big]\,d\eta''}\,e^{iq\eta'}}{\Big[2 \Omega_{k}(\eta')2 \Omega_{p}(\eta')2 q\Big]^{1/2}}\,\, \sum_{a}\overline{\mathcal{U}}_{\vp\,\beta\,a}(\eta')\, {\mathcal{U}_{\vk\,\alpha\,a}}(\eta')\, ,  \label{matxfer1}\ee
 \be \bra{1^{\psi}_{\vk,\alpha}}H_I(\eta)\ket{1^{\psi}_{\vp,\beta};1^{\chi}_{\vq}} = \frac{Y}{V^{1/2}}\,\delta_{\vk,\vp+\vq}\, \frac{e^{i \int^{\eta}_{\eta_i} \big[\Omega_k(\eta'')-\Omega_p(\eta'') \big]\,d\eta''}\,e^{-iq\eta}}{\Big[2 \Omega_{k}(\eta)2 \Omega_{p}(\eta)2 q\Big]^{1/2}}\,\, \sum_{b}\overline{\mathcal{U}}_{\vk\,\alpha\,b}(\eta)\, {\mathcal{U}_{\vp\,\beta\,b}}(\eta)\, \,, \label{matx2fer}\ee with
  \be \Sigma_{k,\alpha}(\eta,\eta') = \sum_{\vp,\vq} \sum_{\beta}\, \bra{1^{\psi}_{\vk,\alpha}}H_I(\eta)\ket{1^{\psi}_{\vp,\beta};1^{\chi}_{\vq}}
 \bra{1^{\psi}_{\vp,\beta};1^{\chi}_{\vq}}H_I(\eta')\ket{1^{\psi}_{\vk,\alpha}}
 \,.\label{sigkfer1}\ee
 Taking the average over the initial polarizations and using the projector (\ref{relaspinors}) we find
 \be \overline{\Sigma}_{k}(\eta,\eta') \equiv \frac{1}{2}\,\sum_{\alpha} \Sigma_{k,\alpha} =
 \frac{Y^2}{16}\,\frac{e^{i \int^{\eta}_{\eta'} \Omega_k(\eta'')\,d\eta''}}{\Big[\Omega_{k}(\eta)\Omega_{k}(\eta')\Big]^{1/2}} \,\, \, \mathcal{I}_k(\eta,\eta')\,, \label{avesigfer}\ee where
  \be \mathcal{I}_k(\eta,\eta') =
 \int \frac{d^3p}{(2\pi)^3}\, \frac{e^{-i \int^{\eta}_{\eta'} \big[\Omega_p(\eta'')+|\vk-\vp| \big]\,d\eta''}}{|\vp-\vk|\,\Big[  \Omega_{p}(\eta)\Omega_{p}(\eta')\Big]^{1/2}}\,
 \mathrm{tr}\Big[\Lambda^{+}_p(\eta,\eta')\,\Lambda^{+}_k(\eta',\eta)\Big]\,.
 \label{Ikerfer}\ee
 Obviously even to leading order in the adiabatic approximation the calculation of the self-energy is a daunting task and no analytic closed expression is available. However, as in the bosonic case of the previous section,   the kernel $\mathcal{I}_k(\eta,\eta')$  is localized in the region $\eta \simeq \eta'$ as a consequence of the momentum integral. Such temporal localization  allows us to leverage the adiabatic expansion to simplify its expression to leading order.

 To understand this aspect more clearly, we follow the same steps as in the bosonic case. In terms of $\epsilon_k(\eta)$ (see eqn. (\ref{epsi})) and $\tau = \Omega_k(\eta)(\eta-\eta')$,   the results (\ref{omex2},\ref{aiden}) lead to the expansion
 \be \mathcal{W}_p(\eta') = \mathcal{W}_p(\eta)\Big[1-\frac{\epsilon_k(\eta)\,\tau}{\gamma_p(\eta)} +\cdots\Big]\,,\label{Wexp}\ee  where the dots stand for higher powers of $\epsilon_k\,\tau$. This identity leads to the expansion
 \be \Lambda^{+}_p(\eta,\eta') = {\Big[ {\not\!{P}}(\eta) + M(\eta)\Big]}+ \frac{\epsilon_k(\eta)\,\tau}{\gamma_p(\eta) }\,\,\widetilde{\Lambda}_p(\eta)\,, \label{adlam}\ee
where $\widetilde{\Lambda}_p(\eta)$ is of zeroth adiabatic order, therefore,
\be  \mathrm{tr}\Big[\Lambda^{+}_p(\eta,\eta')\,\Lambda^{+}_k(\eta',\eta)\Big] = 4\, \Big[ \Omega_k(\eta)\Omega_p(\eta)-\vk\cdot\vp +M^2(\eta)\Big]+ \mathcal{O}(\epsilon_k \tau)\,. \label{traceap}\ee Neglecting the terms of $\mathcal{O}(\epsilon_k \tau)$ the kernel can be written as
\be \mathcal{I}_k(\eta,\eta') = 4\,\int^{\infty}_{-\infty} \widetilde{\rho}(k_0,k)\,e^{-ik_0 T} \,dk_0\,, \label{sperepfer} \ee
where $T$ is the same as for  eqn. (\ref{Ikerap}) and
\be \widetilde{\rho}(k_0,k) =  \int \frac{d^3p}{(2\pi)^3}\,\frac{\delta(k_0-\Omega_p(\eta)-|\vp-\vk|)}{\Omega_p(\eta)\,|\vp-\vk|}\,
\Big[\Omega_p(\eta)\Omega_k(\eta)-\vk\cdot \vp+M^2(\eta) \Big]\,, \label{tilrhofer} \ee which has been calculated in ref.\cite{infrared} and is given by
\be \widetilde{\rho}(k_0,k) =  \frac{1}{8\pi^2 }\, \Bigg[\frac{k^2_0-\Omega^2_k(\eta)}{k^2_0-k^2}\Bigg]\,\Bigg\{k_0\,
 \Bigg[\frac{\Omega_k(\eta)-k_0}{k^2_0-k^2}\Bigg](k^2_0-k^2 +M^2(\eta))+k^2_0-k^2+3M^2(\eta) \Bigg\}   \Theta(k_0-\Omega_k(\eta))   \,. \label{rhofinren}\ee We have suppressed the argument $\eta$ in $\widetilde{\rho}(k_0,k)$ which depends parametrically on it. The short time limit $\eta -\eta' \rightarrow 0$ ($T \rightarrow 0$) is dominated by the large $k_0$ behavior in (\ref{sperepfer}), since for large $k_0$ it follows that $\widetilde{\rho}(k_0,k) \rightarrow k_0 \Omega_k(\eta)$ then as $T\rightarrow 0$
 \be \mathcal{I}_k(\eta,\eta') \propto \frac{1}{(\eta-\eta')^2}\,.  \label{shortifer}\ee The large $T$ behavior is obtained as for the bosonic case, since the spectral density vanishes as $k_0 \rightarrow \Omega_k(\eta)$ following the same steps as for the bosonic case, namely with the identity (\ref{iden}) and the derivative expansion leading to eqn. (\ref{largeT}),  we find the asymptotic long time behavior
 \be \mathcal{I}_k(\eta,\eta') \propto \frac{1}{T^2} \propto \frac{\big(\epsilon_k(\eta)\,\Omega_k(\eta)\big)^2}{\big(\epsilon_k(\eta)\,\tau\big)^2}\,\frac{1}{\Big[1- \overline{\delta}\,\epsilon_k(\eta)\,\tau \Big]^2}\,,  \label{longtifer}\ee therefore for $\tau \simeq 1/\epsilon_k$ when the higher order adiabatic corrections become of the same order as the leading term, the kernel $\mathcal{I}_k$ is of order $\epsilon^2_k$. This analysis leads to the conclusion that the self energy kernel is localized in the region $\eta \simeq \eta'$ and to leading adiabatic order we can set $\eta = \eta'$ in $\Omega_k(\eta'),\Omega_p(\eta')$. Following the same steps as for the bosonic case we find to leading (zeroth) adiabatic order
 \be   \overline{\Sigma}_{k}(\eta,\eta') =  \frac{Y^2}{32\pi^2} \int^{\infty}_{\Omega_k(\eta)}\overline{\rho}(k_0,k)\,e^{-i(k_0-\Omega_k(\eta))(\eta-\eta')}\,dk_0 \,,\label{avesigzero}\ee with
 \be \overline{\rho}(k_0,k) = \frac{(k_0-\Omega_k(\eta))}{\Omega_k(\eta)}\Bigg(\frac{k_0+\Omega_k(\eta)}{k^2_0-k^2} \Bigg)\,\Bigg\{-k_0\,\Bigg[\frac{k_0-\Omega_k(\eta)}{k^2_0-k^2}\Bigg](k^2_0-k^2+M^2(\eta))+k^2_0-k^2+3M^2(\eta)   \Bigg\} \,. \label{overho}\ee We now integrate $\overline{\Sigma}_k$ in $\eta'$ to obtain the decay rate of
 a single fermion with comoving momentum $k$ given by eqn. (\ref{probbaA}), it is given by
\be \Gamma_{k}(\eta)= \frac{Y^2}{16\pi^2}\, \int^\infty_{\Omega_k(\eta)}\,\overline{\rho}(k_0;k)\, \frac{\sin\Big[(k_0-\Omega_k(\eta))(\eta-\eta_i) \Big]}{k_0-\Omega_k(\eta)}\,dk_0\,.  \label{ratafer}\ee
In terms of the variable $s$ defined by eqn. (\ref{svar}) we note that $\overline{\rho}(s)$ given by (\ref{overho})  vanishes linearly in $s$, therefore, we write as for the bosonic case (\ref{tilsigdef}) \be \overline{\rho}(s;\eta) = s\,\mathcal{D}_f(\eta)\,  \Big[1+\,s\,\widetilde{\sigma}_f(s;\eta)  \Big] \,, \label{sigofs}\ee  where for the fermionic case
\be \mathcal{D}_f(\eta)= 8\,\Omega_k(\eta)\,, \label{caldf}\ee and
$\widetilde{\sigma}(s;\eta) \propto s$ as $s\rightarrow 0$. As in the bosonic case, we write $\int^{\infty}_0(\cdots) ds = \int^{1}_{0}(\cdots)ds + \int^{\infty}_{1} (\cdots) ds$, for the first integral   we write $\overline{\rho}(s)$ as in (\ref{sigofs}),   yielding
\be \Gamma_k(\eta) = \Gamma^{(1)}_k(\eta)+
 \Gamma^{(2)}_k(\eta)+\Gamma^{(3)}_k(\eta)\,,\label{gamasfer} \ee with
 \bea \Gamma^{(1)}_k(\eta) & = & \frac{Y^2\,\Omega_k(\eta)}{2\,\pi^2}\,
 \int^1_0 \,\sin\big[s \,\Omega_k(\eta)(\eta-\eta_i)\big]\,ds  \, \,, \label{gama1fer} \\
\Gamma^{(2)}_k(\eta) & = & \frac{Y^2\,\Omega_k(\eta)}{2\,\pi^2}\,
\int^1_0\, s\,\widetilde{\sigma}(s;\eta)\,\sin\big[s \,\Omega_k(\eta)(\eta-\eta_i)\big]\,ds \,, \label{gama2fer} \\
\Gamma^{(3)}_k(\eta) & = & \frac{Y^2}{16\,\pi^2}\,
\int^\infty_1\,   \frac{\overline{\rho}(s;\eta)}{s}\,\sin\big[s \,\Omega_k(\eta)(\eta-\eta_i)\big]\, {ds}   \,, \label{gama3fer} \eea yielding
\be \int^\eta_{\eta_i}\,\Gamma^{(1)}_k(\eta')\,d\eta' = 2\,\Delta_f \, \int^{\eta}_{\eta_i}\,\frac{\Big[1-\cos\Big(\Omega_k(\eta')\,(\eta'-\eta_i) \Big) \Big]}{(\eta'-\eta_i)}\,d\eta'\,,\label{cont1fer} \ee
with
\be \Delta_f = \frac{Y^2}{4\pi^2}\,. \label{deltaf}\ee

The integral for $\Gamma^{(1)}_k(\eta)$ is the same as for the bosonic case, eqn. (\ref{gama1}), therefore the same analysis as that leading to eqn. (\ref{finint}) applies also to (\ref{cont1fer}). An analysis for the contributions from $\Gamma^{(2,3)}_k(\eta)$ is given in appendix (\ref{appendix:gamma23fer}), these yield terms that remain bounded in time at long time but feature ultraviolet divergences. Gathering these terms we find in this case  \be \int^\eta_{\eta_i}\,\Gamma_k(\eta')\,d\eta' = 2\,\Delta_f \,\ln\Big[ \frac{E_k(t)}{H(t)}\Big]+ z_f(t)\,. \label{asyintgamafer}\ee In the fermionic case, $z_f(t)$ is a slowly varying function of $\eta$ that approaches an ultraviolet logarithmically divergent constant in the long time limit. This behavior is manifest in the result given by  (\ref{gama3contfer}) in appendix (\ref{appendix:gamma23fer}) at leading adiabatic order in the long time limit  because the spectral density $\overline{\rho}(s;\eta) \propto s$ for large $s$.
 Therefore, for the fermionic case, the survival probability of a single $\psi$ particle state with momentum $\vk$ in the long time limit is
 \be |C^{\psi}_{\vk,\alpha}(t)|^2 =  |C^{\psi}_{\vk,\alpha}(t_i)|^2\,\Big[ \frac{E_k(t)}{H(t)}\Big]^{-2\Delta_f}\,\mathcal{Z}_f(t) ~~;~~ \mathcal{Z}_f(t) = e^{-z_f(t)}\,, \label{surviprobfer}\ee however in this case the slowly varying wave function renormalization $\mathcal{Z}(t)$ is ultraviolet logarithmically divergent in the long time limit, just as in Minkowski space-time\cite{infrared}.

 Finally, the amplitude of the state $\ket{1^{\psi}_{\vp,\beta};1^{\chi}_{\vq}}$ is
 \be C^{\psi,\chi}_{\vp,\beta;\vk}(\eta)  = -i \int^{\eta}_{\eta_i}  \bra{1^{\psi}_{\vp,\beta};1^{\chi}_{\vq}}H_I(\eta')\ket{1^{\psi}_{\vk,\alpha}}
 \,C^{\psi}_{\vk,\alpha}(\eta')\,d\eta' \ee where the matrix element is given by eqn. (\ref{matxfer1}). Hence, the time evolved state in the interaction picture is
 \be \ket{\Psi_I(\eta)} = C^{\psi}_{\vk,\alpha}(\eta)\,\ket{1^{\psi}_{\vp,\beta};0^{\chi}}+ \sum_{\vp,\beta} C^{\psi,\chi}_{\vp,\beta;\vk}(\eta)\ket{1^{\psi}_{\vp,\beta};1^{\chi}_{\vq}} \,, \label{entferstate}\ee and unitarity (\ref{totder}) implies that
$\langle\Psi_I(\eta)| \Psi_I(\eta)\rangle = |C^{\psi}_{\vk,\alpha}(\eta_i)|^2$.

 \section{Consequences of entanglement:}\label{sec:emt}

 \subsection{Entanglement entropy: information flow.}\label{subsec:ententropy}

 In both, the bosonic and fermionic cases the time evolved states $\ket{\Psi_I(\eta)}$ (\ref{entanbos},\ref{entferstate}) are entangled states of the heavy and the light particle. The pure state density matrix from  $\ket{\Psi_I(\eta)}$ is given by
 \be \widehat{\varrho}(\eta) = \frac{\ket{\Psi_I(\eta)}\bra{\Psi_I(\eta)}}{\langle \Psi_I(\eta)|\Psi_I(\eta)\rangle}\,, \label{purerho} \ee
entanglement is confirmed by obtaining the Von Neumann entanglement entropy from the  reduced density matrix  which is obtained by tracing over one of the degrees of freedom. For example by tracing over the ultralight field $\chi$ for the bosonic case (\ref{entanbos}) we find
 \be \widehat{\varrho}^{\,\varphi}_r(\eta)= \mathrm{Tr}_{\chi} \widehat{\varrho}(\eta) = \big|\widetilde{C}^{\varphi}_k(\eta)\big|^2 \ket{1^{\varphi}_{\vk}}\bra{1^{\varphi}_{\vk}} + \sum_{\vp} \big|\widetilde{C}^{\varphi,\chi}_{\vp;\vk}(\eta)\big|^2\,\ket{1^{\varphi}_{\vp}} \bra{1^{\varphi}_{\vp}} \,\label{redrhophi}\,, \ee  and tracing over the heavy field $\varphi$ we find

 \be \widehat{\varrho}^{\,\chi}_r(\eta)= \mathrm{Tr}_{\varphi} \widehat{\varrho}(\eta) = |\widetilde{C}^{\varphi}_k(\eta)|^2 \ket{0^{\chi}}\bra{0^{\chi}} + \sum_{\vq} \big|\widetilde{C}^{\varphi,\chi}_{\vk-\vq;\vk}(\eta)\big|^2\,\ket{1^{\chi}_{\vq}} \bra{1^{\chi}_{\vq}} \,\label{redrhochi}\,, \ee
  with
 \be \widetilde{C}^{\varphi}_k(\eta) = \frac{ {C}^{\varphi}_k(\eta)}{ {C}^{\varphi}_k(\eta_i)} ~~;~~ \widetilde{C}^{\varphi,\chi}_{\vp;\vk}(\eta) = \frac{ {C}^{\varphi,\chi}_{\vp;\vk}(\eta)}{ {C}^{\varphi}_k(\eta_i)}\,, \label{normalcoefs} \ee it follows from the solutions (\ref{CAfina}) and (\ref{ckapfini}) that the normalized amplitudes $\widetilde{C}^{\varphi}_k(\eta);\widetilde{C}^{\varphi,\chi}_{\vp;\vk}(\eta)$ are \emph{independent} of the initial amplitude ${C}^{\varphi}_k(\eta_i)$ and the unitarity condition (\ref{totder}) yields
 \be \big|\widetilde{C}^{\varphi}_k(\eta)\big|^2+ \sum_{\vp}\big|\widetilde{C}^{\varphi,\chi}_{\vp;\vk}(\eta)\big|^2=1\,,   \label{tildeunit} \ee which implies that
 \be \mathrm{Tr} \widehat{\varrho}^{\,\varphi}_r(\eta)= 1~~;~~ \mathrm{Tr} \widehat{\varrho}^{\,\chi}_r(\eta)= 1 \,. \label{unittraces}\ee

 The reduced density matrices (\ref{redrhophi},\ref{redrhochi})  are diagonal in the basis of single particle states of definite momentum. The von Neumann entropy $S_{vN}(\eta) = - \mathrm{Tr} \widehat{\varrho}_r(\eta)\, \ln(\widehat{\varrho}_r(\eta))$ for both cases is therefore given    by
 \be S_{vN}(\eta) = - \Bigg\{ |\widetilde{C}^{\varphi}_k(\eta)|^2 \,\ln \big[  |\widetilde{C}^{\varphi}_k(\eta)|^2 \big] + \sum_{\vp} |\widetilde{C}^{\varphi,\chi}_{\vp;\vk}(\eta)|^2 \, \ln\big[ |\widetilde{C}^{\varphi,\chi}_{\vp;\vk}(\eta)|^2 \big]  \Bigg\} \,.\label{Svnbose} \ee This entanglement entropy grows during the time evolution since $S_{vN}(\eta_i) =0$ because $\widetilde{C}^{\varphi}_k(\eta_i)=1~;~\widetilde{C}^{\varphi,\chi}_{\vp;\vk}(\eta_i)=0$,    and at very long time
 when the amplitude of the initial state has ``decayed'', namely  $|\widetilde{C}^{\varphi}_k(\eta)|^2=0$ it follows that $S_{vN} > 0$ since $\big|\widetilde{C}^{\varphi,\chi}_{\vp;\vk}(\eta)\big|^2 <1$ as a consequence of the unitarity condition (\ref{tildeunit}) for $|\widetilde{C}^{\varphi}_k(\eta)|^2=0$. The  time evolution  of $S_{vN}$ is completely determined by the (DRM) equations (\ref{eqA},\ref{eqK}) and   describes the information flow from the single particle initial state to the entangled asymptotic final state during the ``dressing'' process.

 \subsection{Energy momentum tensors:}\label{subsec:emts}

The main result of the previous sections is that the amplitude of the initial state
\be \big|\widetilde{C}^{\psi}_{\vk,\alpha}(\eta)\big| \propto \Big[\frac{E_k(\eta)}{H(\eta)}\Big]^{-\Delta_f} ~~;~~ \big|\widetilde{C}^{\varphi}_k(\eta)\big|  \propto  \Big[ \frac{E_k(t)}{H(t)}\Big]^{-\Delta_b}\,.\label{ampsfin}
\ee  To estimate the magnitude of the decay of the amplitude of the initial state between an early period in (RD) to near the radiation to matter transition, let us consider as an example that the mass of the heavy particle   $\simeq \mathrm{GeV}$ and the comoving momentum $k \simeq 10^{-3}\,\mathrm{eV}$ corresponding to an average photon in the cosmic microwave background today. At  the electroweak scale the physical momentum corresponds to $k_{ph}(\eta) \simeq 100\,\mathrm{GeV}$, hence at this scale $E_k(\eta)/H(\eta)\simeq 10^{17}$, whereas near the radiation to matter transition $k_{ph}(\eta) \simeq \mathrm{few}~\mathrm{eV}$ and $E_k(\eta)/H(\eta) \simeq 10^{37}$.

We now study the energy momentum tensor in the asymptotic long time limit, for $\eta \gg \eta_f$ such that the amplitudes of the initial state  $\widetilde{C}^{\varphi}_k(\eta_f); \widetilde{C}^{\psi,\alpha}_k(\eta_f) \simeq 0$ and all the probability in the initial state has flowed to the aymptotic final state with the coefficients $\widetilde{C}^{\varphi,\chi}_{\vp;\vk}(\eta_f);\widetilde{C}^{\psi,\chi}_{\vp;\vk}(\eta_f)$ nearly constant in time and saturating the unitarity relation. In this asymptotic long time limit, the time evolved state is the entangled two particle state $ \ket{\Psi_I(\eta_f)} \simeq  \sum_{\vp} C^{\varphi,\chi}_{\vp;\vk}(\eta_f)\,\ket{1^{\varphi}_{\vp};1^{\chi}_{\vk-\vp}}$ with the coefficients $C^{\varphi,\chi}_{\vp;\vk}(\eta_f)$ being nearly time independent satisfying the unitarity condition
\be  \sum_{\vp} |\widetilde{C}^{\varphi,\chi}_{\vp;\vk}(\eta_f)|^2 \simeq 1  \,,\label{asys} \ee for the bosonic case, with a similar consideration   for the fermionic case.
We are interested in understanding the expectation value of the energy-momentum tensor associated with this state in the asymptotic long time limit $\eta \gg \eta_f$ with $\widetilde{C}^{\varphi}_{\vk}(\eta_f) \simeq 0 ;\widetilde{C}^{\psi}_{\vk}(\eta_f) \simeq 0$, assuming that $\eta_f$ corresponds to a time scale well before recombination.   Let us first consider the bosonic case.

 For minimally coupled fields the energy momentum tensor
 during (RD) (with vanishing Ricci scalar) is\cite{anderson}
 \bea
T_{\mu\nu}(x) & = & \partial_{\mu}\Phi^{\dagger}\partial_{\nu}\Phi+\partial_{\nu}\Phi^{\dagger}\partial_{\mu}\Phi
-g_{\mu\nu}\big[g^{\alpha\beta}\partial_{\alpha}\Phi^{\dagger}\partial_{\beta}\Phi-m^{2}|\Phi|^{2}\big] \nonumber \\
& + &  \partial_{\mu}\pi\partial_{\nu}\pi
-\frac{g_{\mu\nu}}{2}\big[g^{\alpha\beta}\partial_{\alpha}\pi\partial_{\beta}\pi\big] \nonumber \\
& + & \lambda \Phi^\dagger \, \Phi \,\pi \,,
\label{EMT}
\eea covariant conservation can be explicitly confirmed by using the equations of motion\cite{anderson}.

Passing to conformal time and in terms of the conformally rescaled fields (\ref{rescale}) we find \bea
T_{0}^{0}(\vx,\eta)  & = & \frac{1}{a^{4}(\eta)}\Bigg[\Big(\varphi'-\frac{a'(\eta)}{a(\eta)}\varphi\Big)^{\dagger}\,
\Big(\varphi'-
\frac{a'(\eta)}{a(\eta)}\varphi\Big)+
\nabla\varphi^{\dagger}\cdot
\nabla\varphi+M^{2}a^{2}(\eta)\,|\varphi|^{2}\Bigg] \nonumber \\
 & + & \frac{1}{2a^{4}(\eta)}\Bigg[\Big(\chi'-\frac{a'(\eta)}{a(\eta)}\chi\Big)^2 +\nabla\chi\cdot
\nabla\chi + \lambda\, a(\eta)\, \varphi^\dagger\,\varphi\,\chi \Bigg]\,.\label{tzerozero}
 \eea
Upon quantization the energy density becomes an operator in the Heisenberg representation. The energy density of a quantum state $\ket{\Psi}$ is
\be \rho_{\Psi}(\vx,\eta)  = \frac{\langle \Psi| T^0_0 (\vx,\eta)|\Psi\rangle}{\langle \Psi|\Psi\rangle} \,, \label{enerdens}\ee where the state $\ket{\Psi}$ does not evolve in time in the Heisenberg picture.  Since $T^0_0(\vx,\eta) = U^{-1}(\eta,\eta_i)\,(T^0_0(\vx,\eta))_S \,U(\eta,\eta_i)$ where $U(\eta,\eta_i)$ is the time evolution operator (\ref{Uofeta}) and $(T^0_0(\vx,\eta))_S$ is in the Schroedinger picture, where its time dependence is explicit through the scale factor, and writing
as in eqn. (\ref{evolip}) $U(\eta,\eta_i) = U_0(\eta,\eta_i) \,U_I(\eta,\eta_i)$, it follows that
\be \rho_{\Psi}(\vx,\eta)  = \frac{\langle \Psi_I(\eta)|( T^0_0 (\vx,\eta))_I|\Psi_I(\eta)\rangle}{\langle \Psi_I(\eta)|\Psi_I(\eta)\rangle} = \mathrm{Tr}\Big\{  \widehat{\varrho}(\eta) ( T^0_0 (\vx,\eta))_I \Big\}
\,, \label{enerdens2}\ee where $( T^0_0 (\vx,\eta))_I= U^{-1}_{0}(\eta,\eta_i)\,(T^0_0(\vx,\eta))_S \,U_0(\eta,\eta_i)$ is in the interaction picture, wherein the fields carry the free field time evolution (\ref{fige},\ref{chief}). In this form we can now obtain the energy density of the dressed state $\ket{\Psi_I(\eta)}$ given by (\ref{entanbos}) to leading adiabatic order. This is achieved with the following steps: \textbf{i:)} Expand the fields in creation and annihilation operators to leading adiabatic order   as in eqns. (\ref{figezero},\ref{chiefmink}), \textbf{ii:)} neglect the terms with $a'/a, (a'/a)^2$ in $T^0_0$ because these are of first and second adiabatic order respectively, \textbf{iii:)} in the terms quadratic in the fields in $T^0_0$ neglect terms of the
form $a^\dagger b^\dagger, a b, c^\dagger c^\dagger, c c$ because the asymptotic state $\ket{\Psi_I}$ contains terms of the form $\ket{1^\varphi}\ket{1^\chi}$ namely products of single particle states for each particle, hence expectation values of the form $\langle \Psi_I|a^\dagger b^\dagger |\Psi_I\rangle =0$ and similarly with the other bilinears, \textbf{iv:)} the expectation value of the interaction term $\langle \Psi_I|\varphi^\dagger \varphi \chi |\Psi_I\rangle =0$,  because $\chi \simeq c + c^\dagger$ hence either destroying or creating a single $\chi$ particle
from $\ket{\Psi_I}$ therefore the expectation value of such operator vanishes. As a result the expectation value of the energy momentum tensor becomes a sum of the contribution from the heavy field $\varphi$ and that of the ultra light field $\chi$. For each of these, the expectation value implies tracing over the other field (for example for the contribution of the $\varphi$ field, it implies tracing over the $\chi$ field, and viceversa). Therefore  we find that asymptotically at long time, when the amplitude of the initial single particle state has become negligible, and to leading order in the coupling
\be \rho_{\Psi}(\eta) = \frac{1}{a^4(\eta)\,V}\,\sum_{\vp} \Omega_{p}(\eta)\,\mathrm{Tr}\Big\{ \widehat{\varrho}^{\,\varphi}_r(\eta)\big(a^\dagger_{\vp} a_{\vp}+ b^\dagger_{\vp} b_{\vp} +1\big)\Big\}   + \frac{1}{a^4(\eta)\,V}\,\sum_{\vq}|\vq|\, \mathrm{Tr}\Big\{ \widehat{\varrho}^{\,\chi}_r(\eta) \big(c^\dagger_{\vp} c_{\vp}  +\frac{1}{2}\big) \Big\} \,,\label{finT00}\ee  where $V$ is the comoving volume, and $\widehat{\varrho}^{\,\varphi}_r(\eta)~;~\widehat{\varrho}^{\,\chi}_r(\eta) $ are the reduced density matrices (\ref{redrhophi},\ref{redrhochi}) respectively.

The terms $(1,1/2)$ inside the respective parenthesis in (\ref{finT00}) yield the zero point energy which as usual is subtracted away with an appropriate renormalization scheme (this is usually assumed in the literature), and we find
\be \rho_{\Psi}(\eta) = \frac{1}{a^3(\eta)} \int \frac{d^3p}{(2\pi^3)}\,E_p(\eta)\, \big| \widetilde{C}^{\varphi,\chi}_{\vp;\vk}(\eta_f)\big|^2 + \frac{1}{a^4(\eta)} \int \frac{d^3q}{(2\pi^3)}\,|q|\,\, \big| \widetilde{C}^{\varphi,\chi}_{\vk-\vq;\vk}(\eta_f)\big|^2\equiv \rho^{M}(\eta)+ \rho^{R}(\eta)\,.  \label{enedensphi}\ee  Asymptotically at long time when the single particle amplitude of the heavy field has ``decayed'',  $ \big|\,\widetilde{C}^{\varphi,\chi}_{\vp;\vk}(\eta_f)\big|^2$ becomes a \emph{non-thermal frozen distribution function} fulfilling the ``sum rule'' (\ref{asys}) from the unitarity condition in the asymptotic long time limit. The energy density (\ref{enedensphi}) describes two independent fluids: the first term, $\rho^{M}(\eta)$ is identified with the energy density of a massive, frozen species, and the second $\rho^{R}(\eta)$ with a massless,  frozen ultrarelativistic species, both independently obeying covariant conservation, namely
\be \frac{d}{dt} \rho^{M}(t)+ 3 H(t) \big(\rho^{M}(t)+\mathcal{P}^{M}(t)\big) =0 ~~;~~ \mathcal{P}^{M}(t)  = \frac{1}{3} \int \frac{d^3p_{ph}}{(2\pi)^3}\frac{p^2_{ph}}{E_p(t)}\,
\big| \widetilde{C}^{\varphi,\chi}_{\vp;\vk}(\eta_f)\big|^2\,,\label{pressure}\ee

\be \frac{d}{dt} \rho^{R}(t)+  4 H(t)  \rho^{R}(t)  =0  \,.\label{covaconsrad} \ee

The expression (\ref{finT00}) for the expectation value of the energy density involves the reduced density matrices $\widehat{\varrho}^{\,\varphi}_r(\eta); \widehat{\varrho}^{\,\chi}_r(\eta)$ obtained by tracing over the $\chi,\varphi$ fields respectively. This suggests that the entropy associated with each fluid is precisely the entanglement entropy (\ref{Svnbose}),   because each fluid component in the energy momentum tensor arises from tracing over the complementary field yielding the reduced density matrices (\ref{redrhophi},\ref{redrhochi}) each of which describes a \emph{mixed state} associated with the entanglement entropy (\ref{Svnbose}). Entanglement in the final asymptotic state entails that the fluids share the \emph{same} entropy and the same frozen distribution function.

It is important to highlight that we have studied the time evolution of an initial \emph{single particle state}, as a result the energy density and pressure are both proportional to $1/V$ since the matrix elements yielding the coefficients $\widetilde{C} \propto 1/V$ (see for example eqns. (\ref{matx1})), therefore at long time the unitarity condition (\ref{asys}) yields
\be \int \frac{d^3p}{(2\pi^3)} \big| \widetilde{C}^{\varphi,\chi}_{\vp;\vk}(\eta_f)\big|^2 = \frac{1}{V}\,, \label{singpart} \ee this is the statement that there is one $\varphi$ and also one $\chi$ particle in the volume $V$ in the final state. We discuss this aspect in section (\ref{sec:discussion}).

For the case of fermionic fields Yukawa coupled to the ultralight scalar field, using the field equations for the Dirac field\cite{birrell},  the energy momentum tensor is given by  \cite{parkerbook,fercurved3,feradia,feradia2,feradia3}
 \be T^{\mu \nu} = \frac{i}{2} \Big( \overline{\Psi} \gamma^\mu  \stackrel{\leftrightarrow}{\mathcal{D}^\nu}   \,\Psi \Big) + \mu \leftrightarrow \nu +  \partial^{\mu}\pi\partial^{\nu}\pi
- \frac{1}{2}\,{g^{\mu\nu}}   {g^{\alpha\beta}} \partial_{\alpha}\pi\partial_{\beta}\pi
   \label{tmunudirac} \ee

 In terms of conformal time and the conformally rescaled fields (\ref{rescaledfields}) and using again the field  equations for the Dirac field\cite{birrell} to restore the Yukawa interaction term,
 the energy density $T^0_0$ is given by
 \be    T^0_0(\vx,\eta)  = \frac{1}{ a^4(\eta)} \Bigg\{      \,\psi^\dagger (\vec{x},\eta) \Big(-i\vec{\alpha}\cdot \vec{\nabla} + \gamma^0 M\,a(\eta)   \Big)\psi (\vec{x},\eta)  +  \frac{1}{2} \Big(\chi'-\frac{a'(\eta)}{a(\eta)}\chi\Big)^2 +\frac{1}{2}\,\nabla\chi\cdot
\nabla\chi +  Y \,\psi^\dagger \chi \psi \Bigg\} \,. \label{rhoopfer}\ee

As in the bosonic case, we pass to the interaction picture and obtain the energy density corresponding to the
time evolved state $\ket{\Psi_I(\eta)}$ now given by eqn. (\ref{entferstate}) as in eqn. (\ref{enerdens2}) and follow
the same steps as in the bosonic case. Again, considering a long time $\eta_f$ after which the amplitude of the initial single fermion state has ``decayed'', the time evolved state is given by $\sum_{\vp,\beta} C^{\psi,\chi}_{\vp,\beta;\vk}(\eta_f)\ket{1^{\psi}_{\vp,\beta};1^{\chi}_{\vq}}$, hence to leading order in the coupling the expectation value of the Yukawa interaction term in this asymptotic state vanishes because in the interaction picture the field $\chi \simeq c + c^\dagger$ whose expectation value vanishes in this state.
The fermion fields in the interaction picture are expanded as in eqn. (\ref{psiex}) where the spinors are the solutions of the Dirac equations (\ref{Uspinor},\ref{Vspinor})  with normalization given by eqn. (\ref{normas}). To leading adiabatic order they are given by (\ref{Uspinorsol},\ref{Vspinorsol}) and obey $\partial_{\eta} U_{\lambda}(\vk,\eta) = \Omega_k(\eta)\,U_{\lambda}(\vk,\eta);\partial_{\eta} V_{\lambda}(\vk,\eta) = -\Omega_k(\eta)\,V_{\lambda}(\vk,\eta)$. Since the expectation value of the Yukawa interaction in the interaction picture vanishes in the asymptotic state,  to leading order order in Yukawa coupling and adiabatic expansion  the energy density associated with this asymptotic state is a sum of the free fermion and free bosonic fields energy densities. In turn these contributions are determined by the corresponding \emph{reduced} density matrices.  For  the fermionic term we need
the reduced density matrix $\widehat{\varrho}^{\,\psi}_r(\eta)= \mathrm{Tr}_{\chi} \widehat{\varrho}(\eta)$ obtained by tracing the $\chi $ degrees of freedom, whereas the bosonic one inputs the reduced density matrix $\widehat{\varrho}^{\,\chi}_r(\eta)= \mathrm{Tr}_{\psi} \widehat{\varrho}(\eta)$ obtained by tracing over the fermionic degree of freedom. We finally find that the energy density associated with the asymptotic state is given by
\be \rho_{\Psi}(\eta) = \frac{1}{a^4(\eta)\,V}\,\sum_{\vp,\lambda} \Omega_{p}(\eta)\,\mathrm{Tr}\Big\{ \widehat{\varrho}^{\,\psi}_r(\eta)\big(b^\dagger_{\vp,\lambda} b_{\vp,\lambda}+ d^\dagger_{\vp,\lambda} d_{\vp,\lambda} +1\big)\Big\}   + \frac{1}{a^4(\eta)\,V}\,\sum_{\vq}|\vq|\, \mathrm{Tr}\Big\{ \widehat{\varrho}^{\,\chi}_r(\eta) \big(c^\dagger_{\vp} c_{\vp}  +\frac{1}{2}\big) \Big\} \,,\label{finT00fer}\ee
Just as in the bosonic case, the terms $(1,1/2)$ inside the   parenthesis yield the zero point energy which  is subtracted away with an appropriate renormalization scheme  yielding
\be \rho_{\Psi}(\eta) = \frac{1}{a^3(\eta)} \int \frac{d^3p}{(2\pi^3)}\,E_p(\eta)\, \big| \widetilde{C}^{\psi,\chi}_{\vp;\vk}(\eta_f)\big|^2 + \frac{1}{a^4(\eta)} \int \frac{d^3q}{(2\pi^3)}\,|q|\,\, \big| \widetilde{C}^{\psi,\chi}_{\vk-\vq;\vk}(\eta_f)\big|^2\equiv \rho^{M}(\eta)+ \rho^{R}(\eta)\,.  \label{enedensfer}\ee  Asymptotically at long time when the single particle amplitude of the heavy field has ``decayed'', and $ \big|\,\widetilde{C}^{\psi,\chi}_{\vp;\vk}(\eta_f)\big|^2$ becomes a \emph{non-thermal frozen distribution function}. The energy density (\ref{enedensfer}) again describes two independent fluids:   $\rho^{M}(\eta)$ is identified with the energy density of a massive, fermionic non-thermal frozen species, and the second $\rho^{R}(\eta)$ with a massless, ultrarelativistic non-thermal frozen species, both obeying covariant conservation, as in the bosonic case (\ref{pressure},\ref{covaconsrad}) but with $|\widetilde{C}^{\varphi,\chi}_{\vk-\vq;\vk}(\eta_f)\big|^2 \rightarrow |\widetilde{C}^{\psi,\chi}_{\vk-\vq;\vk}(\eta_f)\big|^2$. Both fluids share the same frozen distribution function $|\widetilde{C}^{\psi,\chi}_{\vk-\vq;\vk}(\eta_f)\big|^2$  and entanglement entropy,
 \be S_{vN}(\eta) = - \Bigg\{  \sum_{\vp} |\widetilde{C}^{\psi,\chi}_{\vp;\vk}(\eta_f)|^2 \, \ln\big[ |\widetilde{C}^{\psi,\chi}_{\vp;\vk}(\eta_f)|^2 \big]  \Bigg\} \,,\label{Svnfer} \ee
since as in the bosonic case, each component in the energy momentum tensor emerges from tracing the complementary field.

It is noteworthy that the entanglement entropy of the asymptotic state from infrared dressing is \emph{very different} from from that of cosmological particle production which leads to a  squeezed state\cite{entropydm,beilokentropy}.

 \section{Discussions:}\label{sec:discussion}

 \textbf{On gravitational particle production:}

 Gravitational particle production is negligible in the cases that we have considered in this study for the following reasons.  The adiabatic approximation relies on the mass of the heavy field being much larger than the Hubble expansion rate, the terms in the (time dependent) Hamiltonian that would yield gravitational production are of second or higher order in the adiabatic expansion, therefore subleading. This is explicit in the  terms with $a_{\vk}\,b_{-\vk};b^\dagger_{-\vk}\, a^\dagger_{\vk}$ in the Hamiltonian for the bosonic case, eqn. (\ref{Hphi}). These terms would lead to particle production but they are multiplied by a function which is of second or higher adiabatic order\cite{decaycosmo,scattering} which can be neglected to the leading adiabatic order implemented in this study. Furthermore,   we considered the light scalar field to be (nearly) massless, and a massless scalar field is conformally coupled to gravity in a radiation dominated cosmology because the Ricci scalar vanishes. Therefore there is not gravitational production of the  light scalar field either during the radiation era.

 \vspace{1mm}

 \textbf{Dressing vs. Decay:}

  Consider the case of two massive fields $\phi^{(1,2)}$ with masses $M_1 > M_2$, and a massless field $\chi$ with a coupling $\lambda \phi^{(1)}\,\phi^{(2)}\,\chi$ and the decay process  $\phi^{(1)}_{\vk}\rightarrow \phi^{(2)}_{\vp}+\chi_{\vq}$.  At time much longer than the lifetime of $\phi^{(1)}$ the asymptotic final state is given by $\sum_{\vq} \mathcal{C}_{\vq,\vk} \, \ket{1^{\phi}_{\vk-\vq};1^{\chi}_{\vq}}$, this is kinematically entangled two particle state and unitarity leads to $\sum_{\vq} |\mathcal{C}_{\vq,\vk}|^2 = |C^{1}_k(t_i)|^2$ where $C^{1}_k(t_i)$ is the amplitude of the single particle initial state\cite{boyww}.  This state is qualitatively similar to (\ref{entanbos}) asymptotically when $C^{\varphi}_k(\eta) \simeq 0$. The only differences are: \textbf{a:)} in particle decay the amplitude of the single particle state decays exponentially but with a decay law modified by the cosmological expansion\cite{decaycosmo}, whereas for infrared dressing it decays with a power law with anomalous dimension, \textbf{b:)} in the case of decay, the final two-particle state does not contain the initial particle, whereas in the case of infrared dressing, the initial massive particle is part of the entangled final state. These differences notwithstanding, particle  decay leads to the production of daughter particles in a kinematically entangled  final state.
  The expectation value of the energy momentum tensor in the asymptotic final state will feature independent contributions from the daughter particles with   negligible contribution from the interaction term because the final state does not contain the particle in the initial state. Again final state entanglement implies that both contributions have the same frozen distribution function. Hence  the analogy with the final asymptotic state from infrared dressing is compelling and indicates that this latter mechanism also leads to the production of the massless particle in the final state. This interpretation is confirmed by the expectation value of the energy momentum tensor in the asymptotic state obtained in the previous section.  The important aspect is that in both cases the amplitude of the initial state vanishes at long time and by unitarity, the total probability flows entirely from the initial state to the final entangled state. Furthermore, in both cases, entanglement in the asymptotic state implies that the daughter particles share the frozen distribution and entanglement entropy.

 \vspace{1mm}

 \textbf{Dressing of entangled pairs:}

 In this article we focused on studying the time evolution of an initial single particle state and obtained the time evolved state to leading order in the adiabatic and weak coupling approximations. However, we did not specify the mechanism by which the initial state has been prepared.  Heavy massive particles can be produced gravitationally prior to the radiation era, however these are described by an entangled squeezed state (see for example refs.\cite{herringdm,entropydm,beilokentropy} and references therein) not as independent single particle states. Squeezed states are highly correlated, and it is an open question, relegated to future study, whether pair correlations modify the dynamics of infrared dressing, and if so how the pair correlations in the initial state are manifest in the asymptotic entangled state.

 \vspace{1mm}

 \textbf{Single particle vs. density matrix:}

 We have focused on studying the dynamics of infrared dressing for a single heavy particle. As a result the  distribution function for the asymptotic state is   given by eqn.(\ref{singpart}), namely $\propto 1/V$, indicating that in the final state there is only one massless and one massive particle. Therefore, although the  fundamental  study of infrared dressing in the single particle case provides a ``proof of principle'' of a mechanism of  production of ultra light dark matter or dark radiation, obviously it is not very cosmologically relevant \emph{yet} because a cosmologically relevant dark matter or radiation candidate   requires a finite {density} in the infinite volume limit. The next step is to consider an ensemble of heavy particles described by a density matrix in terms of a distribution function for the heavy degrees of freedom. The time evolution of such density matrix  would be determined by a Boltzmann-like equation that should follow from the dynamical resummation method implemented in this study. This next step in the program will be the focus of forthcoming studies.

\vspace{1mm}

 \textbf{Distribution function of ULDM?:}

 In Minkowski space time the results of ref.\cite{infrared} showed that the pair probability or distribution function for the bosonic case of the asymptotic entangled state is $|\widetilde{C}^{\varphi,\chi}_{\vk,\vq}(\infty)|^2\propto [E_k+q-E_p]^{2\Delta-2}$, with a similar result for the fermionic case.    Although we did not calculate it explicitly in the cosmological case, based on the similarities between the cosmological result and that in Minkowski space time at leading adiabatic order, we expect a similar result for the distribution to leading adiabatic order with the energies replaced by the local energies depending on the scale factor at a time scale when the amplitude of the initial state becomes negligible. Although this expectation is motivated by the results obtained in the previous sections and the similarity with Minkowski space-time at leading adiabatic order, it must be confirmed by a detailed analysis. Such calculation is technically involved and neither very illuminating nor relevant for the question of dark matter because it is associated with an initial \emph{single particle state}, hence its contribution to the energy momentum tensor is $\propto 1/V$ (see eqn. (\ref{singpart})) hence negligible   in the infinite volume limit, and not relevant to dark matter. Our goal with this study is to provide a ``proof of principle'' of infrared dressing as a possible production mechanism and to pave the way towards a future study of an initial state described by a finite density  ensemble of heavy fields described by an appropriate density matrix. Undoubtedly the asymptotic distribution function obtained from the time evolution of this density matrix will reflect the finite density aspects of the initial distribution yielding a finite contribution to the energy momentum tensor in the infinite volume limit. This will be the subject of a forthcoming study, which is now motivated by this ``proof of principle''.

 \vspace{1mm}

 \textbf{On axions:}  The study of ref.\cite{infrared} in Minkowski space time revealed that in the case of fermions coupled to pseudoscalar fields, such as the axion, the spectral density vanishes faster than linear at threshold. As a result these type of couplings do not yield infrared divergences in Minkowski space-time. In this case the amplitude of the initial single particle case does not vanish asymptotically and the unitarity condition is satisfied at long time with the amplitude of the initial state being nearly the same as that at the initial time with a perturbatively small probability for axion production from infrared dressing.

 This result discouraged a similar study in cosmology suggesting that infrared dressing may not be an important mechanism of production of axions during the radiation era. Nevertheless a derivative type coupling  such as $g_{\mu \nu} \partial^{\mu} \mathcal{A}(x)\,\overline{\Psi}(x) \gamma^\nu(x)\,\gamma^5\,\Psi(x)$ with $\mathcal{A}(x)$ the pseudoscalar field, \emph{may}   lead to some interesting phenomena which, however we postpone to further study.

 \vspace{1mm}

 \textbf{Radiative corrections to ultralight mass:} Masses of scalar or pseudoscalar fields are in general subject to large radiative corrections unless there are symmetries that lead to their cancellations. Otherwise the small values are the result of some fine tuning. Ultra-light scalar particles as originally envisaged in the form of ``fuzzy'' dark matter\cite{fuzzyDM,fuzzy2,wittenfuzzy,hui} would be subject to (divergent) radiative corrections if not protected by a symmetry as for example (pseudo) Goldstone bosons. Therefore the question of radiative corrections in principle apply to generic ``fuzzy'' dark matter models.  In our study, focused on the fundamental aspects of infrared dynamics,  we have simply assumed that the (nearly) massless scalar degree of freedom remains (nearly) massless after radiative corrections. Therefore, the application of our results to any phenomenological extension beyond the Standard Model must assess whether the (near) masslessness of this ultra light dark matter or dark radiation candidate remains robust under radiative corrections.

 \vspace{1mm}

 \textbf{Caveats: very weak couplings.} There is an important caveat in the results obtained in the previous section, namely we assumed that the amplitude of the initial state becomes negligible during (RD) (or the early stages in the matter dominated era). However, unlike particle decay where the amplitude of the initial state decays (nearly) exponentially\cite{decaycosmo}, infrared dressing yields to a power law decay, which is much slower. The anomalous dimension $\Delta$ in the decay law (\ref{surviprob}) is proportional to the square of the coupling, hence very small for very weak coupling. Therefore, it is possible that for very weak couplings, the amplitude of the initial state remains substantial near the end of the (RD) era and the contribution of the initial state dominates the energy momentum tensor, and only later during the matter era  the ultra light or dark radiation component begins to contribute appreciably to the relativistic component of the energy momentum tensor. If the heavy bosonic or fermionic species are suitable dark matter candidates, this scenario introduces the possibility of a dark radiation component to be produced during the matter era. Clearly these possibilities must be studied in detail within a phenomenologically
 viable model, which goes well beyond the scope of this initial study.

 \section{Conclusions and further questions}\label{sec:conclusions}

 The main objectives of this article are to study the fundamental aspects of infrared phenomena in a radiation dominated cosmology,   and to provide a ``proof of principle'' of infrared dressing as a hitherto unexplored possible production mechanism of ultra light dark matter or dark radiation. Infrared dressing describes the cloud of massless quanta that dresses the heavy particle as a consequence of emission and absorption of nearly on shell massless quanta. Infrared aspects of these processes are ubiquitous  in gauge theories and in gravity arising from the emission and absorption of massless gauge bosons or gravitons.

 We focused on a bosonic and a fermionic theory of heavy fields coupled to a nearly massless   scalar field as prototypes of non-gauge  quantum field theories featuring infrared divergences.

 We combined an adiabatic approximation valid for wavelengths much smaller than the Hubble radius with a non-perturbative dynamical resummation method to study the time evolution of an initial single particle state. This method is manifestly unitary and consistently describes the time evolved state.

 The massless (or nearly) massless scalar field may be associated with an ultra light dark matter or dark radiation candidate in extensions beyond the Standard Model. However, we are neither proposing nor endorsing particular phenomenological extensions beyond the Standard Model, focusing solely on the fundamental aspects of infrared dynamics and their possible cosmological consequences.

 We showed that as a result of   infrared divergences the amplitude of the initial single particle state decays in time with a power law $\propto [E_k(t)\,t]^{-\Delta}$ with $E_k(t)$ being the local energy depending on the scale factor as a consequence of the cosmological redshift, entailing a crossover from $t^{-\Delta/2}$ during the relativistic regime to $t^{-\Delta}$ upon becoming non-relativistic. This decay law is common to bosonic or fermionic degrees of freedom suggesting certain universality for infrared phenomena in cosmology. The anomalous dimension $\Delta$ is determined by the behavior of the spectral density near threshold. The quantum state that emerges in the asymptotic long time limit after the initial state has decayed is an entangled state of the heavy boson or fermion and the massless scalar, with amplitudes that are completely determined by unitary time evolution and yield the \emph{ frozen distribution function} of the final state.

 Quantum entanglement is confirmed by obtaining the von Neumann entanglement entropy by tracing either degree of freedom. The time evolution of the entanglement entropy is completely determined by the dynamical resummation equations, it increases during time evolution   and describes the flow of information from the initial single particle to the asymptotic entangled many particle states.

 We argued that  infrared dressing as a production mechanism is qualitatively similar to that of particle decay in that the amplitude of the initial state vanishes at long time and the asymptotic state is an entangled state of the daughter particles. The mayor difference is that in the decay process the initial amplitude vanishes exponentially (or nearly exponentially in an expanding cosmology\cite{decaycosmo}) rather than with a power law with anomalous dimension as is the case of infrared dressing.

 To leading order in the adiabatic expansion and in weak coupling, the expectation value of the energy momentum tensor in the asymptotic state describes two independent fluids one associated with the heavy boson or fermion and another associated with a relativistic degree of freedom, namely either the ultra light dark matter or dark radiation. Both fluids fulfill the covariant conservation equation independently. An important consequence of entanglement in the asymptotic state is that both fluids share the same non-thermal frozen distribution function and entropy.

 Gathering these results together this study suggests that infrared dressing is a possible production mechanism of ultra light dark matter and or dark radiation with basic features that are qualitatively similar to production via particle decay. Because we have considered a simple initial state and the study provides a ``proof of principle'' of the fundamental and ubiquitous phenomenon of infrared dressing as a viable production mechanism, many questions remain that merit further and deeper study. Among them the extrapolation of the single particle case to that of an ensemble of heavy degrees of freedom coupled to (nearly) massless scalars and, in particular if this ensemble is a result of gravitational production of the heavy degrees of freedom with a particular distribution function. We also recognized important caveats in the case of very weak couplings. Furthermore, while discouraged by the results in Minkowski space time\cite{infrared}, whether a pseudovector coupling in a cosmological setting yields to interesting infrared phenomena remains an open question. This study also paves the way towards understanding of infrared phenomena associated with massless gauge bosons or gravitons. However the issue of gauge invariance and concomitant fulfillment of Ward identities during the dynamical evolution remains to be understood for a consistent treatment.  The possibility that this mechanism may contribute to the understanding of dark matter or dark radiation production thus motivates further studies along these avenues.

\appendix

\section{Contributions from $\Gamma^{(2,3)}_k(\eta)$ for the bosonic case.}\label{appendix:gamma23}
With the definition (\ref{Deltadef}), the change of variables (\ref{xdefs}) and taking $\eta \gg \eta_i$ the contribution from $\Gamma^{(2)}_k$ yields
\be \int^{\eta}_{\eta_i} \,\Gamma^{(2)}_k(\eta') d\eta' = 2\,\Delta_b \, \int^{1/\epsilon_k(\eta)}_0 \,dx \int^1_0 \, ds \,\widetilde{\sigma}(s;\eta') \, \frac{s\, \sin[s\,x]}{1+\widetilde{\epsilon}_k(x) x}\,, \label{2ndcon} \ee
where where $\eta'$ depends implicitly on $x$ via (\ref{xdefs}).
Writing $\sin[sx] = -\frac{1}{s} \,d \cos[sx]/dx$ and integrating by parts in $x$,  the integral in (\ref{2ndcon}) becomes

\bea  && \underbrace{\int^1_0 ds \Bigg\{ \widetilde{\sigma}(s;\eta_i)- \frac{\widetilde{\sigma}(s;\eta)}{1+\frac{1}{\gamma^2_k(\eta)}}\, \cos[s/\epsilon_k(\eta)] \Bigg\}}_{A} \nonumber \\ + && \underbrace{\int^{1/\epsilon_k(\eta)}_0 dx \int^1_0 ds \frac{d}{dx} \Bigg\{\frac{\widetilde{\sigma}(s;\eta')}{(1+\widetilde{\epsilon}_k(\eta')\,x)}  \Bigg\}\,\cos[sx] }_{B}\,.
\eea
In the first term (A) the cosine term averages out in the long time limit $\epsilon_k(\eta) \rightarrow  0$. Using the following identities:
\be \frac{d\,\widetilde{\epsilon}_k(\eta')}{dx} = \frac{d\,\widetilde{\epsilon}_k(\eta')}{d\eta'}\,\frac{d\eta'}{dx}= \frac{d}{d\eta'}\Big(\frac{\Omega'_k(\eta')}{\Omega^2_k(\eta') }\Big)\,\frac{1}{\Omega_k(\eta')\,(1+\widetilde{\epsilon}_k(\eta')\,x)}\,, \ee

\be \frac{1}{\Omega_k(\eta')}\,\frac{d}{d\eta'}\Big(\frac{\Omega'_k(\eta')}{\Omega^2_k(\eta') }\Big) = \frac{\Omega^{''}_k(\eta')}{\Omega^3}-2\,\Big(\frac{\Omega'_k(\eta')}{\Omega^2_k(\eta') }\Big)^2\,, \ee this latter term being second order adiabatic,  and
\be \frac{d}{dx} \Big( \frac{1}{\gamma^2_k(\eta')}\Big)= \frac{d}{d\eta'} \Big( \frac{1}{\gamma^2_k(\eta')}\Big)\,\frac{1}{\Omega_k(\eta')\,(1+\widetilde{\epsilon}_k(\eta')\,x)}
\,,\ee along with the identity
\be \frac{1}{\Omega_k(\eta')}\, \frac{d}{d\eta'} \Big( \frac{1}{\gamma^2_k(\eta')}\Big)= \frac{2\,\epsilon_k(\eta')}{\gamma^2_k(\eta')}\Big[1-\frac{1}{\gamma^2_k(\eta')} \Big]\,,\ee
using all these identities, we find that the contribution (B) is infrared and ultraviolet finite and at least of first adiabatic order ($\mathcal{O}(\epsilon_k)$), hence it can be safely neglected to leading adiabatic order. Therefore, to leading adiabatic order we find
\be \int^{\eta}_{\eta_i} \,\Gamma^{(2)}_k(\eta') d\eta' = 2\,\Delta_b \, \int^1_0    \widetilde{\sigma}(s,\eta_i)\, ds  \,. \label{contgama2fina}\ee
Following the same steps for the third contribution, we find
\be \int^{\eta}_{\eta_i} \,\Gamma^{(3)}_k(\eta') d\eta' = \frac{\lambda^2}{4} \int^{1/\epsilon_k(\eta)}_0 dx \,\int^\infty_1 ds \, \frac{\overline{\rho}(s,\eta')}{s} \, \frac{a^2(\eta')}{\Omega^2_k(\eta')}\, \frac{\sin[sx]}{(1+\widetilde{\epsilon}_k(\eta')\,x)}\,.\label{3rdcont}  \ee
  Implementing the same steps for the integrals as for the second contribution  yields the following result for the integrals in (\ref{3rdcont})
\bea && \underbrace{\int^\infty_1 \Bigg\{ {\overline{\rho}(s,\eta_i)} )\,\frac{a^2(\eta_i)}{\Omega^2_k(\eta_i)}-
  {\overline{\rho}(s,\eta)} \,\frac{a^2(\eta)}{\Omega^2_k(\eta)}\,\frac{\cos[s/\epsilon_k(\eta)]}
  {1+\frac{1}{\gamma^2_k(\eta)}}\Bigg\}\,\frac{ds}{s^2}}_{A} \nonumber \\ + && \underbrace{\int^{1/\epsilon_k(\eta)}_0 dx \int^\infty_1 \frac{ds}{s^2} \frac{d}{dx} \Bigg\{\frac{\overline{\rho}(s,\eta')\,a^2(\eta')}{\Omega^2_k(\eta')\,(1+\widetilde{\epsilon}_k(\eta')\,x)}  \Bigg\}\,\cos[sx] }_{B}\,. \eea The oscillatory cosine term in (A) averages out in the long time limit, and implementing the same steps and definitions as for the second contribution, the (B) term is found to be both infrared and ultraviolet finite and of (at least) first adiabatic order, hence subleading and averaging out in the long time limit.
 Therefore, we find in the long time limit  $\Omega_k(\eta)\eta \gg 1$ ($\eta >> \eta_i$) and to leading adiabatic order
 \be \int^{\eta}_{\eta_i} \,\Gamma^{(3)}_k(\eta') d\eta'  = \frac{\Delta_b}{\gamma^2_k(\eta_i)}\,\int^\infty_1 \frac{2+s}{\frac{1}{\gamma^2_k(\eta_i)} +2s+s^2 } \,\frac{ds}{s} \, \label{gama3cont} \ee which is an infrared and ultraviolet finite constant.

 \section{Contributions from $\Gamma^{(2,3)}_k(\eta)$ for the fermionic case.}\label{appendix:gamma23fer} For the fermionic case $\Gamma_k^{(2,3)}$ are given respectively by eqns. (\ref{gama2fer},\ref{gama3fer}). With the definition (\ref{deltaf}) and the change of variables (\ref{xdefs}) we find
 \be \int^{\eta}_{\eta_i} \,\Gamma^{(2)}_k(\eta') d\eta' = 2\,\Delta_f \, \int^{1/\epsilon_k(\eta)}_0 \,dx \int^1_0 \, ds \,\widetilde{\sigma}_f(s;\eta') \, \frac{s\, \sin[s\,x]}{1+\widetilde{\epsilon}_k(x) x}\,. \label{2ndconfer} \ee Implementing the same steps as for the bosonic case in the previous section, we find to leading adiabatic order and at long time
 \be \int^{\eta}_{\eta_i} \,\Gamma^{(2)}_k(\eta') d\eta' = 2\,\Delta_f \, \int^1_0    \widetilde{\sigma}_f(s,\eta_i)\, ds  \,, \label{contgama2finafer}\ee
 which is infrared and ultraviolet finite and
 \be \int^{\eta}_{\eta_i} \,\Gamma^{(3)}_k(\eta') d\eta'  =   \frac{{\Delta_f}}{4} \,\int^\infty_1 \frac{\overline{\rho}(s;\eta_i)}{\Omega_k(\eta_i)} \,\frac{ds}{s^2} \,, \label{gama3contfer} \ee which however diverges logarithmically because $\overline{\rho}(s;\eta) \propto s$ as $s\rightarrow \infty$, reflecting the renormalizability of the Yukawa coupling to a scalar field.

\acknowledgements
D. B. gratefully acknowledges support from the U.S. National Science Foundation through award 2111743.

\end{document}